\documentclass[a4paper,fleqn,usenatbib]{mnras}
\usepackage{longtable}
\usepackage{newtxtext,newtxmath}
\usepackage[T1]{fontenc}
\usepackage{ae,aecompl}
\usepackage{graphicx}	
\defcitealias{ngc5904}{Paper I}
\defcitealias{ngc6205}{Paper II}
\defcitealias{ngc288}{Paper III}
\defcitealias{ngc6362}{Paper IV}
\defcitealias{nardiello2018}{NLP18}
\defcitealias{simioni2018}{SBA18}
\defcitealias{stetson2019}{SPZ19}
\defcitealias{grundahl1999}{GCL99}
\defcitealias{twarog2000}{AT2000}
\defcitealias{mandushev1998}{MFR}
\defcitealias{vasiliev2021}{VB21}
\defcitealias{ccm89}{CCM89}
\defcitealias{bonatto2013}{BCK13}
\title[Isochrone fitting of NGC\,6397 and NGC\,6809]
{Isochrone fitting of Galactic globular clusters -- V. NGC\,6397 and NGC\,6809 (M55)}

\author[G. A. Gontcharov et al.]{
George~A.~Gontcharov,$^{1}$\thanks{E-mail: georgegontcharov@yahoo.com}
Charles~J.~Bonatto,$^{2}$ 
Olga~S.~Ryutina,$^{3}$
\newauthor
Sergey~S.~Savchenko,$^{1,3,4}$
Aleksandr~V.~Mosenkov,$^{5,1}$
Vladimir~B.~Il'in,$^{1,3,6}$
\newauthor
Maxim~Yu.~Khovritchev,$^{1,3}$
Alexander~A.~Marchuk,$^{1,3}$
Denis~M.~Poliakov,$^{1,3}$
\newauthor
Anton~A.~Smirnov,$^{1,3}$
and Jonah~Seguine$^{5}$
\\
$^{1}$Central (Pulkovo) Astronomical Observatory, Russian Academy of Sciences, Pulkovskoye chaussee 65/1, St. Petersburg 196140, Russia\\
$^{2}$Departamento de Astronomia, Instituto de F\'isica, UFRGS, Av. Bento Gon\c{c}alves, 9500, Porto Alegre, RS, Brazil\\
$^{3}$Saint Petersburg State University, Universitetskij pr. 28, St. Petersburg 198504, Russia\\
$^{4}$Special Astrophysical Observatory, Russian Academy of Sciences, 369167 Nizhnij Arkhyz, Russia\\
$^{5}$Department of Physics and Astronomy, Brigham Young University, N283 ESC, Provo, UT 84602, USA\\
$^{6}$Saint Petersburg University of Aerospace Instrumentation, Bol. Morskaya ul. 67A, St. Petersburg 190000, Russia\\
}

\date{Accepted 2023 September 28. Received 2023 September 28; in original form 2023 January 31}

\pubyear{2023}

\begin{document}
\label{firstpage}
\pagerange{\pageref{firstpage}--\pageref{lastpage}}
\maketitle

\begin{abstract}
We fit various colour--magnitude diagrams (CMDs) of the Galactic globular clusters NGC\,6397 and NGC\,6809 (M55) by isochrones from the Dartmouth Stellar Evolution Database (DSED) 
and Bag of Stellar Tracks and Isochrones (BaSTI) for $\alpha$--enhanced [$\alpha$/Fe]$=+0.4$.
For the CMDs, we use data sets from {\it HST}, {\it Gaia}, VISTA, and other sources utilizing 32 and 23 photometric filters for NGC\,6397 and NGC\,6809, respectively, 
from the ultraviolet to mid-infrared.
We obtain the following characteristics
 for NGC\,6397 and NGC\,6809, respectively:
metallicities [Fe/H]$=-1.84\pm0.02\pm0.1$ and $-1.78\pm0.02\pm0.1$ (statistic and systematic uncertainties);
distances $2.45\pm0.02\pm0.06$ and $5.24\pm0.02\pm0.18$ kpc; 
ages $12.9\pm0.1\pm0.8$ and $13.0\pm0.1\pm0.8$ Gyr;
reddenings $E(B-V)=0.178\pm0.006\pm0.01$ and $0.118\pm0.004\pm0.01$ mag; 
extinctions $A_\mathrm{V}=0.59\pm0.01\pm0.02$ and $0.37\pm0.01\pm0.04$ mag;
extinction-to-reddening ratio $R_\mathrm{V}=3.32^{+0.32}_{-0.28}$ and $3.16^{+0.66}_{-0.56}$.
Our estimates agree with most estimates from the literature.
BaSTI gives systematically higher [Fe/H] and lower reddenings than DSED. 
Despite nearly the same metallicity, age, and helium enrichment, these clusters show a considerable horizontal branch (HB) morphology difference, which must 
therefore be described by another parameter.
This parameter must predominantly explain why the least massive HB stars (0.58--0.63 solar masses) are only found within NGC 6809.
Probably they have been lost by the core-collapse cluster NGC\,6397 during its dynamical evolution and mass segregation.
In contrast, NGC\,6809 has a very low central concentration and, hence, did not undergo this process.
\end{abstract}


\begin{table*}
\def\baselinestretch{1}\normalsize\small
\caption[]{Reddening converted into $E(B-V)$ (mag) and age (Gyr) estimates for NGC\,6397 or NGC\,6809 from recent isochrone fittings of CMDs.
}
\label{previous}
\[
\begin{tabular}{lllcccc}
\hline
\noalign{\smallskip}
Study                     &  Isochrones    &  Data set and colour                        & \multicolumn{2}{c}{NGC\,6397} & \multicolumn{2}{c}{NGC\,6809} \\
\hline
\noalign{\smallskip}
                          &                &                                             &    $E(B-V)$    &    Age       &   $E(B-V)$    &   Age         \\
\hline
\noalign{\smallskip}
\citet{dotter2010}        & DSED           & {\it HST}/ACS, $F606W-F814W$                &      0.18      & $13.5\pm0.5$ &     0.11      &  $13.5\pm1.0$  \\
\citet{dicriscienzo2010}  & own            & {\it HST}/ACS, $F606W-F814W$                &                & $13.0\pm1.0$ &               & \\
\citet{siegel2011}        & DSED           & {\it HST}/ACS, $F606W-F814W$                &                &              & $0.12\pm0.01$ &      \\
\citet{vandenberg2013}    & VR             & {\it HST}/ACS, $F606W-F814W$                &                & $13.0\pm0.3$ &               & $13.0\pm0.3$ \\
\citet{martinazzi2014}    & DSED           & dedicated ground-based, $B-V$               &  $0.12\pm0.01$ & $12.0\pm0.5$ &               &  \\
\citet{chen2014}          & PARSEC         & {\it HST}/ACS, F606W-F814W                  &      0.20      & 12.0         &               & \\
\citet{campos2016}        & own            & {\it HST}/ACS, $F606W-F814W$                &  $0.18\pm0.01$ & $12.9\pm0.4$ &               &   \\ 
\citet{correnti2018}      & VR             & {\it HST}/WFC3, IR $F110W-F160W$            &  $0.22\pm0.02$ & $12.6\pm0.7$ &               &   \\
\citet{tailo2020}         & own            & {\it HST}/WFC3/ACS, $F438W-F814W$           &                & $13.0\pm0.5$ &               & $13.3\pm0.5$ \\
\citet{valcin2020}        & DSED           & {\it HST}/ACS, $F606W-F814W$                &  $0.15\pm0.01$ & $14.2\pm0.7$ & $0.08\pm0.01$ & $13.9\pm0.6$  \\
\citet{ahumada2021}       & VR             & dedicated ground-based, $V-I$               &       0.19     & $13.3\pm0.3$ &               &   \\
\hline
\end{tabular}
\]
\end{table*}

\section{Introduction}
\label{intro}

In \citet[][hereafter Paper I]{ngc5904}, \citet[][hereafter Paper II]{ngc6205}, \citet[][hereafter Paper III]{ngc288}, and \citet[][hereafter Paper IV]{ngc6362} 
we used theoretical stellar evolution models and their corresponding  isochrones to fit colour--magnitude diagrams (CMDs) for the Galactic globular clusters (GCs)
NGC\,288, NGC\,362, NGC\,5904 (M5), NGC\,6205 (M13), NGC\,6218 (M12), NGC\,6362, and NGC\,6723.

This series of papers is inspired by the recent appearances and improvements for models/isochrones and photometric data sets of individual cluster members in 
ultraviolet (UV), optical, and infrared (IR) bands.
In particular, we use photometric data from the {\it Hubble Space Telescope (HST}; \citealt{piotto2015}, 
\citealt[][hereafter NLP18]{nardiello2018}, \citealt[][hereafter SBA18]{simioni2018}), 
{\it Gaia} Data Release 2 (DR2; \citealt{evans2018}), Early Data Release 3 (EDR3; \citealt{riello2021}), and Data Release 3 (DR3; \citealt{gaiadr3}), 
{\it Wide-field Infrared Survey Explorer (WISE}; \citealt{wise}) as the unWISE catalogue \citep{unwise}, various ground-based telescopes by \citet[][hereafter SPZ19]{stetson2019}, 
and other sources.
Moreover, the precise parallaxes and proper motions (PMs) from {\it HST} and {\it Gaia} EDR3\footnote{The photometry and astrometry of GCs are exactly the same in {\it Gaia} EDR3 and DR3.} 
allow us an accurate selection of GC members.
To fit CMDs we use theoretical models of stellar evolution, namely
the Dartmouth Stellar Evolution Database (DSED, \citealt{dotter2007})\footnote{\url{http://stellar.dartmouth.edu/models/}} and
a Bag of Stellar Tracks and Isochrones (BaSTI, \citealt{pietrinferni2021})\footnote{\url{http://basti-iac.oa-abruzzo.inaf.it/index.html}}.
These models are presented by user-friendly online tools in order
to calculate isochrones for low metallicity, various levels of helium abundance, and $\alpha$--enhancement, which are typical in GCs \citep{monelli2013,milone2017}.
These isochrones reproduce different stages of stellar evolution, namely the main sequence (MS), turn-off (TO), subgiant branch (SGB), red giant branch (RGB), 
horizontal branch (HB), and asymptotic giant branch (AGB).
Best-fitting isochrones provide us with age, distance, reddening, and metallicity [Fe/H] for a cluster dominant population or a mix of populations.

We cross-identify data sets to estimate systematic differences between them, convert the derived reddenings into extinction for each filter we consider, and draw 
an empirical extinction law (i.e. a dependence of extinction on wavelength) for each combination of cluster, data set, and model.

In this paper, we fit the pair of GCs NGC\,6397 and NGC\,6809 (also known as Messier 55 or M55). 
These clusters are considerably contaminated by foreground and background stars (in particular, those of Sagittarius dwarf galaxy, see \citealt{siegel2011}).
Hence, their data sets should be cleaned with PMs and parallaxes. 
Accordingly, we can fit isochrones directly to a bulk of certain cluster members in very clean CMDs, without calculation of any fiducial sequence.
Since these clusters are similar in metallicity, age, helium enrichment, and reddening, it is fruitful to consider their relative estimates.
Moreover, this similarity makes NGC\,6397 and NGC\,6809 
interesting to find an explanation for their significant HB morphology difference besides metallicity, age, and helium enrichment.

Previous isochrone fittings of the clusters since \citet{dotter2010} and corresponding reddening and age values are presented in Table~\ref{previous}.
Their results can be compared with ours (see Sect.~\ref{results}).
Unfortunately, earlier fittings by \citet{alcaino1997,twarog2000,gratton2003,richer2008} for NGC\,6397 and by \citet{piotto1999} for NGC\,6809 
seem to be obsolete due to incomplete or too simple models.

NGC\,6397 and NGC\,6809 are among metal-poor GCs, with [Fe/H]<-1.7. They are valuable candidates to verify modern models/isochrones in such a low-metallicity regime, 
especially in application to the data sets never fitted before.

Most of the studies in Table~\ref{previous} fit Victoria-Regina (VR, \citealt{vandenberg2018}) or DSED isochrones to the {\it HST}/ACS data sets.\footnote{The solar-scaled 
PAdova-TRieste Stellar Evolution Code (PARSEC, \citealt{bressan2012}) isochrones, used by \citet{chen2014}, seem to be inappropriate for GCs.} 
Our study stands out, since it engages the BaSTI isochrones (together with the DSED ones) and many more data sets, which have appeared or improved recently.
The number of data sets and photometric measurements, fitted in our study, is an order of magnitude higher than in any study before.

The outline of this paper is as follows.
We present some properties of NGC\,6397 and NGC\,6809 in Sect.~\ref{clusters}, theoretical models and isochrones used -- in Sect.~\ref{iso}, 
and data sets used -- in Sect.~\ref{photo}.
The results of our isochrone fitting are introduced and discussed in Sect.~\ref{results}.
In Sect.~\ref{conclusions}, we summarize our main findings and conclusions.

\begin{table*}
\def\baselinestretch{1}\normalsize\normalsize
\caption[]{Some properties of the clusters under consideration.
}
\label{properties}
\[
\begin{tabular}{lcc}
\hline
\noalign{\smallskip}
 Property            &  NGC\,6397  &  NGC\,6809 (M55) \\
\hline
\noalign{\smallskip}
RA J2000 (h~m~s) from \citet{goldsbury2010}                                                     & \hphantom{$-$}17 40 42     &     \hphantom{$-$}19 40 00  \\
Dec. J2000 ($\degr$ $\arcmin$ $\arcsec$) from \citet{goldsbury2010}                             & $-53$ 40 28                &     $-30$ 57 53              \\
Galactic longitude ($\degr$) from \citet{goldsbury2010}                                         & 338.1650                   &     8.7926                \\
Galactic latitude ($\degr$) from \citet{goldsbury2010}                                          & $-11.9595$                 &     $-23.2716$              \\
Tidal radius (arcmin) from \citet{moreno2014}                                                   & 44.5                       & 15.3                   \\
Angular radius (arcmin) from \citet{bica2019}                                                   &   11.5                     & 19.0           \\
Truncation radius (arcmin) from this study                                                      & 41.0                       & 18.0                \\
Distance from the Sun (kpc) from \citet{harris}, 2010 revision\footnotemark\                    & 2.3                        &     5.4                      \\
Distance from the Sun (kpc) from \citet{baumgardt2021}                                          &   $2.482\pm0.019$          &     $5.348\pm0.052$            \\
$[$Fe$/$H$]$ from \citet{carretta2009}                                                          & $-1.99\pm0.02$             &  $-1.93\pm0.02$              \\
$[$Fe$/$H$]$ from \citet{meszaros2020}                                                          & $-1.89\pm0.09$             &  $-1.76\pm0.07$              \\
$[\alpha/$Fe$]$ from \cite{carretta2010}                                                        & $+0.36$                    & $+0.42$                    \\
Mean differential reddening $\overline{\Delta E(B-V)}$ (mag) from 
\citetalias{bonatto2013}      & $0.019\pm0.009$            &   $0.027\pm0.010$  \\
Maximum differential reddening $\Delta E(B-V)_\mathrm{max}$ (mag) from 
\citetalias{bonatto2013} & 0.051                      &   0.050            \\
$E(B-V)$ (mag) from \citet{harris}, 2010 revision                                               & 0.18     & 0.08 \\
$E(B-V)$ (mag) from \citet{sfd98}                                                               & 0.19     & 0.14 \\ 
$E(B-V)$ (mag) from \citet{schlaflyfinkbeiner2011}                                              & 0.16     & 0.12 \\ 
$E(B-V)$ (mag) from \citet{planck}                                                              & 0.31     & 0.15 \\
$E(B-V)$ (mag) from \citet{gms2022}                                                             & 0.17     & 0.12  \\
\hline
\end{tabular}
\]
\end{table*}
\footnotetext{The commonly used database of GCs by \citet{harris} (\url{https://www.physics.mcmaster.ca/~harris/mwgc.dat}), 2010 revision.}

\section{Properties of the clusters}
\label{clusters}

Table~\ref{properties} presents some properties of NGC\,6397 and NGC\,6809.

Two or even three populations are known in the clusters \citep{dicriscienzo2010,vandenberg2018}.
All the populations of both the clusters are $\alpha$--enriched with [$\alpha$/Fe]$\approx0.4$ \citep{carretta2010,rain2019,meszaros2020}.

The populations of each cluster differ in helium abundance $Y$. 
\citet{milone2017} estimated the fraction of the first (primordial) population of stars as $0.345\pm0.036$ in NGC\,6397 and $0.311\pm0.029$ in NGC\,6809.
\citet{vandenberg2018} found the fractions of three populations in NGC\,6809 as 51\%, 41\%, and 8\% with $Y=0.25$, 0.265, and 0.28, respectively.
\citet{mucciarelli2014} derived a nearly primordial average $Y=0.24\pm0.02$ for NGC\,6397 using a large data set of helium abundances obtained with a high-resolution spectrograph.
\citet{milone2018} found a small average helium difference between the populations, as well as a small maximum internal helium variation: 
$\Delta Y=0.006\pm0.009$ and $\Delta Y_{max}=0.008\pm0.011$ for NGC\,6397 and $\Delta Y=0.014\pm0.008$ and $\Delta Y_{max}=0.026\pm0.015$ for NGC\,6809.
\citet{lagioia2021} found no evidence of intrinsic broadening of the AGB due to helium abundance variation for both NGC\,6397 and NGC\,6809.
However, such broadening of the RGB is seen in some of our CMDs and may suggest that NGC\,6809 has some stars with rather high $Y$, as discussed in Sect.~\ref{fitting}.
\citet{kaluzny2014} derived $Y\approx0.25$ for NGC\,6809 from their consistent mass-radius, mass-luminosity, and CMD fitting of the cluster's eclipsing binary V54.
We verify in Sect.~\ref{results} that $Y\approx0.26$ also does not contradict to the properties of V54.
In Sect.~\ref{iso}, we take this information into account to select appropriate $Y$ for our isochrone-to-data fitting.

Table~\ref{properties} demonstrates that both NGC\,6397 and NGC\,6809 have rather accurate metallicity estimates from spectroscopy by \citet{carretta2009}.
However, later [Fe/H] estimates demonstrate some issues.
For example, for NGC\,6809, \citet{rain2019} found very low [Fe/H]$=-2.01\pm0.02$ from their analysis of UVES spectra of 11 stars, while \citet{wang2017} found [Fe/H]$=-1.86\pm0.06$
using the same technique applied to UVES and GIRAFFE spectra.
Moreover, \citet{meszaros2020} derived average [Fe/H]$=-1.89\pm0.09$ and $-1.76\pm0.07$ for NGC\,6397 and NGC\,6809, respectively, from high-resolution spectra of a hundred RGB 
stars of each cluster.
\citet{meszaros2020} found that their [Fe/H] are about 0.15 dex systematically higher than those from \citet{carretta2009} and from the compilation of \citet{harris}, not only for 
NGC\,6397 and NGC\,6809, but for the bulk of GCs in their sample. They noted that most of the [Fe/H] difference can be explained through the choice of the reference solar abundance 
mixture,\footnote{\citet{meszaros2020} and \citet{carretta2009} are based on the solar metallicity estimates from \citet{grevesse2007} and \citet{gratton2003}, respectively. 
\citet{gm2018} show that models of the Galaxy combined with stellar evolution models make solar metallicity tightly related to estimates of reddening/extinction across the whole 
Galactic dust layer above or below the Sun. Namely, the higher the reddening (more dusty Galaxy) the lower the solar metallicity.
In combination with the discussion of \citet{meszaros2020}, this shows a relation between [Fe/H] of GCs and the amount of dust in our Galaxy.}
while the remaining difference may be due to wrong calibrations because of wrong reddening estimates or due to a systematic difference in the temperature scales used, or due to 
some effects which are not modeled yet.
Besides, [Fe/H] estimates from photometry are not always consistent with each other and do not always agree with the estimates of \citet{carretta2009}: 
e.g. \citet{correnti2018} found [Fe/H]$=-1.88\pm0.04$ for NGC\,6397 from isochrone fitting of IR photometry of the faint MS.
Moreover, \citet{lovisi2012} used high-resolution spectra of NGC\,6397 stars at various stages and found [Fe/H]$=-2.12\pm0.01$ for the TO, while $-1.20\pm0.22$ for blue stragglers 
and $-1.94\pm0.14$ for the HB.
To explain such discrepancies, \citet{jain2020} analyze a variation of metallicity between the TO and RGB through the use of hundreds stellar spectra and conclude that at the 
low metallicity regime of [Fe/H]$\approx-2$ both synthetic and empirical stellar spectra need to be improved against a considerable systematics.
Also these discrepancies may be due to diffusive processes combined with convection affecting in different ways the stars in distinct evolutionary stages \citep{cassisi2020}.
Thus, we see that the current absolute accuracy of the iron scale seems to be no higher than $\pm0.1$ dex and, hence, needs improvement and verification by different methods.

Photometry can provide [Fe/H] estimates for such a verification. The slopes of the RGB and faint MS ($>3$ mag fainter than TO), are sensitive to 
[Fe/H].\footnote{Note that isochrones show large systematic errors in the faint MS domain. Hence, we consider this domain with caution (see Sect.~\ref{metal}).}
We get [Fe/H] as an isochrone fitting parameter (along with reddening, age, and distance) in CMDs with well-populated bright RGB or faint MS. 
The average values of the derived [Fe/H] estimates are used for fitting the remaining CMDs. We consider it separately for each model.

However, both the bright RGB and faint MS are affected by helium enrichment \citep{savino2018}, crowding or poor astrometry at the cluster field centres, saturation and 
completeness effects, and systematic errors of photometry.
These effects may result in a systematic uncertainty of about 0.2 dex in our [Fe/H] estimate obtained from the pair of a CMD and a model.

Table~\ref{properties} indicates a relatively small foreground and differential reddening for NGC\,6397 and NGC\,6809.
However, the reddening estimates in Table~\ref{properties} are not fully consistent when taking into account their stated precision as a few hundredths of a magnitude, 
which is confirmed by \citet{gm2018}.
Namely, the outliers are a very high reddening estimate by \citet{planck} for NGC\,6397 and a very low estimate by \citet{harris} for NGC\,6809.
Interestingly, in \citetalias{ngc6362}, the reddening estimate by \citet{planck} is the higher outlier for NGC\,6362, which is located in the same fourth Galactic quadrant as 
NGC\,6397, while the reddening estimate by \citet{harris} is the lower outlier for NGC\,6723, which is located in the same first Galactic quadrant as NGC\,6809.
This may be a result of spatial variations of dust medium properties.

Among the reddening estimates in Table~\ref{properties}, a pair is taken from the new version of our 3D analytical model of dust spatial distribution \citep{gms2022}, 
whose predictions agree with isochrone-to-CMD reddening estimates for most middle- and high-latitude GCs.

A mild differential reddening across the field of these clusters is shown by \citet{alonso2012,milone2012} and \citet[][hereafter BCK13]{bonatto2013}.
We correct data sets with a large number of cluster members for differential reddening in Sect.~\ref{difred}.

\section{Theoretical models and isochrones}
\label{iso}

We use the following theoretical stellar evolution models and corresponding $\alpha$--enhanced isochrones to fit the CMDs of NGC\,6397 and NGC\,6809:
\begin{enumerate}
\item BaSTI \citep{newbasti,pietrinferni2021} with various [Fe/H] and helium abundance, [$\alpha$/Fe]$=+0.4$, initial solar $Z_{\sun}=0.0172$ and $Y_{\sun}=0.2695$, 
diffusion, overshooting, mass loss efficiency $\eta=0.3$, where $\eta$ is the free parameter in Reimers law \citep{reimers}.
As in \citetalias{ngc288} and in \citetalias{ngc6362}, we also apply the BaSTI extended set of zero-age horizontal branch (ZAHB) models with the same mass for the helium 
core and the same envelope chemical stratification but different values for the total mass.
This set seems to be a realistic description of stochastic mass loss between the MS and HB.
\item DSED \citep{dotter2008} with various [Fe/H] and helium abundance, [$\alpha$/Fe]$=+0.4$, solar $Z_{\sun}=0.0189$ and no mass loss. 
DSED gives no realistic ZAHB with a stochastic mass loss taken into account. However, DSED provides the HB and AGB isochrones for some filters.
We do not use these isochrones for our fitting, following a recommendation by the DSED team \citep[][private communication]{dotter}.
Yet, we present these isochrones in some our CMD figures in order to show that an acceptable description of the HB and AGB is possible with DSED too. 
\end{enumerate}

In order to fix an appropriate $Y$ for isochrone fitting of each cluster, we fit isochrones with different $Y$ to all CMDs under consideration and conclude that 
BaSTI isochrones with $Y>0.25$ better fit the blue AGB domain, but only for NGC\,6809 and only for data sets covering the whole cluster field, while both BaSTI and DSED 
isochrones with $Y>0.25$ better fit the faint RGB domain in all CMDs.
\text{As noted in \citetalias{ngc6362}, this widening of the faint RGB may be due to}
 a segregation into two populations, with higher and lower helium abundance.
The remaining CMD domains are fitted by the $Y=0.267$ and 0.25 isochrones equally well, since these isochrones almost coincide.
Thus, for our isochrone-to-CMD fitting, we adopt $Y=0.25$ for an unresolved mix of the populations in both the clusters, except the blue AGB and faint RGB, which are fitted with
$Y=0.267$ (see Sect.~\ref{fitting}). 
Only the isochrones with $Y=0.25$ are shown in our CMD figures for clarity.

We fit the isochrones for a grid within $-2.4<$[Fe/H]$<-1.5$ with a step of 0.1 dex, 
distances within $\pm0.8$ kpc from the \citet{baumgardt2021} estimates with a step of 0.01 kpc, 
reddenings between zero and twice the highest reddening estimate from Table~\ref{properties} with a step of 0.001 mag, 
and ages within $8-18$ and $8-15$ Gyr for BaSTI and DSED, respectively, with a step of 0.5 Gyr.

\section{Data sets}
\label{photo}

\subsection{Initial data sets}
\label{datasets}

The following data sets (hereafter {\it twin} data sets, see Table~\ref{filters}) are used for both the clusters:
\begin{enumerate}
\item the {\it HST} Wide Field Camera 3 (WFC3) UV Legacy Survey of Galactic Globular Clusters (the $F275W$, $F336W$, and $F438W$ filters) and the Wide Field Channel of the Advanced Camera 
for Surveys (ACS; the $F606W$ and $F814W$ filters) survey of Galactic globular clusters \citep{piotto2015}, 
\citepalias{nardiello2018},\footnote{\url{http://groups.dfa.unipd.it/ESPG/treasury.php}}
with additional photometry of the same RGB, SGB, and MS stars of NGC\,6397 in the WFC3 $F467M$, ACS $F435W$, and ACS $F625W$ filters  
\citep{libralato2022},\footnote{\url{https://archive.stsci.edu/hlsp/hacks}}
\item Parallel-Field Catalogues (the ACS $F475W$ and $F814W$ filters) of the {\it HST} UV Legacy Survey of Galactic Globular Clusters 
\citepalias{simioni2018},\footnote{\url{http://groups.dfa.unipd.it/ESPG/treasury.php}}
\item $UBVRI$ photometry from various ground-based telescopes processed by \citetalias{stetson2019},\footnote{\url{http://cdsarc.u-strasbg.fr/viz-bin/cat/J/MNRAS/485/3042}} with the NGC\,6397 and NGC\,6809 data sets 
processed within the same pipeline and presented recently,\footnote{\url{https://www.canfar.net/storage/vault/list/STETSON/homogeneous/Latest_photometry_for_targets_with_at_least_BVI}}
\item {\it Gaia} DR3 photometry in the $G$, $G_\mathrm{BP}$ and $G_\mathrm{RP}$ filters \citep{riello2021},\footnote{In \citetalias{ngc6362} we checked that the DSED isochrones 
for DR2 are equally suitable for DR3 and, hence, they are shown in our CMDs with the DR3 data.}
\item SkyMapper Southern Sky Survey DR3 (SMSS, SMSS DR3) photometry in the $v_\mathrm{SMSS}$, $g_\mathrm{SMSS}$, $r_\mathrm{SMSS}$, $i_\mathrm{SMSS}$, and $z_\mathrm{SMSS}$ filters 
\citep{onken2019},\footnote{\url{https://skymapper.anu.edu.au}}
\item $J_\mathrm{VISTA}$ and $Ks_\mathrm{VISTA}$ photometry of the VISTA Hemisphere Survey with the VIRCAM instrument on the Visible and Infrared Survey Telescope for Astronomy 
(VISTA,VHS DR5; \citep{vista}),\footnote{\url{https://cdsarc.cds.unistra.fr/viz-bin/ReadMe/II/367?format=html&tex=true};
DSED does not provide isochrones for VISTA filters. However, as in \citetalias{ngc6362}, $J_\mathrm{VISTA}$ is substituted by $J_\mathrm{UKIDSS}$, a filter from the 
United Kingdom Infrared Telescope Infrared Deep Sky Survey (UKIDSS; \citealt{ukidss}), with a precision better than 0.01 mag.}
\item {\it Wide-field Infrared Survey Explorer (WISE)} photometry in the $W1$ filter from the unWISE catalogue 
\citep{unwise}.\footnote{\url{https://cdsarc.cds.unistra.fr/viz-bin/cat/II/363}}
\end{enumerate}

The following data sets are used for one of the clusters (see Table~\ref{filters}):
\begin{enumerate}
\item {\it HST}/ACS photometry of NGC\,6397 in the $F606W$ and $F814W$ filters \citep{richer2008},\footnote{\url{https://cdsarc.cds.unistra.fr/viz-bin/cat/J/AJ/135/2141}}
\item photometry of NGC\,6397 in the $F439W$ and $F555W$ filters from the {\it HST} Wide Field and Planetary Camera 2 (WFPC2) \citep{piotto2002},\footnote{\url{http://groups.dfa.unipd.it/ESPG/hstphot.html}}
\item Str\"omgren $uvby$ photometry of NGC\,6397 with the 1.54-m Danish Telescope, European Southern Observatory (ESO), La Silla \citep[][hereafter GCL99]{grundahl1999},
\item CCD Str\"omgren $uvby$ and $V$ photometry of NGC\,6397 with the 0.9-m telescope at Cerro-Tololo Inter-American Observatory (CTIO) 
\citep[][hereafter AT2000]{twarog2000},\footnote{\url{https://cdsarc.cds.unistra.fr/viz-bin/cat/J/AJ/120/3111}}
\item $BVI$ photometry of NGC\,6397 with the FOcal Reducer and low dispersion Spectrograph 2 (FORS2) mounted at the Very Large Telescope (VLT) UT1 of the European Southern 
Observatory (ESO) \citep{nardiello2015},\footnote{\url{http://groups.dfa.unipd.it/ESPG/followup.html}. We do not use the photometry in the $U$ filter, since the BaSTI and DSED isochrones
cannot fit its faint MS with reliable parameters, while \citet{nardiello2015} provide few stars with the $U$ photometry in the remaining CMD domains.}
\item $BV$ photometry of NGC\,6397 with the 0.9-m CTIO telescope \citep{kaluzny1997},\footnote{\url{https://cdsarc.cds.unistra.fr/viz-bin/cat/J/A+AS/122/1}}
\item $VI$ photometry of NGC\,6397 with the 1.54-metre telescope of the Bosque Alegre Astrophysical Station of the C\'ordoba Observatory, National University of C\'ordoba, 
Argentina and 1-metre Swope Telescope of Las Campanas Observatory, Chile \citep{ahumada2021},
\item the Two Micron All Sky Survey (2MASS) photometry of NGC\,6397 in the $J_\mathrm{2MASS}$ filter \citep{2mass},
\item $BV$ photometry of NGC\,6809 with the 2.5-m du Pont telescope at Las Campanas Observatory \citep{kaluzny2010},\footnote{\url{https://case.camk.edu.pl/results/Photometry/M55/index.html}}
\item $BV$ photometry of NGC\,6809 with the 1-m Swope telescope of Las Campanas Observatory \citep{narloch2017},\footnote{\url{https://cdsarc.cds.unistra.fr/viz-bin/cat/J/MNRAS/471/1446}}
\item the fiducial sequences for NGC\,6809 in the $BVI$ filters derived by \citet{mandushev1996} and \citet[][hereafter MFR]{mandushev1998} from the photometry with the 2.5-m Du Pont 
telescope of the Las Campanas Observatory.
\item the Panoramic Survey Telescope and Rapid Response System Data Release I (Pan-STARRS, PS1, \citealt{chambers2016}) photometry of NGC\,6809 in the 
$g_\mathrm{PS1}$ and $i_\mathrm{PS1}$ filters.\footnote{NGC\,6809 is near the declination limit of PS1 at about $-30\degr$ and, hence, we use only photometry in 
$g_\mathrm{PS1}$ and $i_\mathrm{PS1}$, which has the best quality.}
\end{enumerate}

Most of these data sets have never been isochrone-fitted before, since they appeared recently.

Three data sets with the {\it HST}/ACS photometry, i.e. presented by (i) \citet{richer2008}, (ii) \citetalias{simioni2018}, and (iii) \citetalias{nardiello2018} together with
the photometry from \citet{libralato2022} for some of the same stars, have little, if any, common stars. 
The Str\"omgren photometry data sets of \citetalias{grundahl1999} and \citetalias{twarog2000} are independent.
All of the $BVI$ data sets from \citetalias{stetson2019}, \citet{nardiello2015}, \citet{kaluzny1997}, \citet{kaluzny2010}, \citet{narloch2017}, and \citetalias{mandushev1998} are independent.
The \citetalias{stetson2019} data sets contain photometry from various initial data sets, but not from the others under consideration.

\begin{table*}
\def\baselinestretch{1}\normalsize\small
\caption[]{The effective wavelength $\lambda_\mathrm{eff}$ (nm), number of stars, and the median precision of the photometry (mag) for the data sets and filters under consideration.
For data sets cross-identified with {\it Gaia} DR3, only {\it Gaia} cluster members are counted.
}
\label{filters}
\[
\begin{tabular}{llccc}
\hline
\noalign{\smallskip}
 Telescope, data set, reference & Filter & $\lambda_\mathrm{eff}$ &  \multicolumn{2}{c}{Number of stars~/~Median precision} \\
\hline
\noalign{\smallskip}
           &        &                        & NGC\,6397 & NGC\,6809 \\
\hline
\noalign{\smallskip}
{\it HST}/WFC3 \citepalias{nardiello2018}                         & $F275W$             & 285  & 5093 / 0.01 & 7007 / 0.01   \\
{\it HST}/WFC3 \citepalias{nardiello2018}                         & $F336W$             & 340  & 6514 / 0.01 & 9212 / 0.01   \\
1.54-m Danish telescope, ESO, La Silla \citepalias{grundahl1999}  & Str\"omgren $u$     & 349  & 5207 / 0.02 & -- \\
0.9-m CTIO telescope \citepalias{twarog2000}                      & Str\"omgren $u$     & 349  & 2466 / 0.02 & -- \\
Various \citepalias{stetson2019}                                  & $U$                 & 366  & 11403 / 0.02 & 7529 / 0.02   \\
SkyMapper Sky Survey DR3 \citep{onken2019}                        & $v_\mathrm{SMSS}$   & 385  & 3494 / 0.02 & -- \\
1.54-m Danish telescope, ESO, La Silla \citepalias{grundahl1999}  & Str\"omgren $v$     & 414  & 5619 / 0.01 & -- \\
0.9-m CTIO telescope \citepalias{twarog2000}                      & Str\"omgren $v$     & 414  & 2470 / 0.01 & -- \\
{\it HST}/ACS \citep{libralato2022}                               & $F435W$             & 434  & 11678 / 0.01 & -- \\
{\it HST}/WFC3 \citepalias{nardiello2018}                         & $F438W$             & 438  & 7339 / 0.01 & 11207 / 0.01    \\
{\it HST}/WFPC2 \citep{piotto2002}                                & $F439W$             & 452  & 5407 / 0.01 & --    \\
Various \citepalias{stetson2019}                                  & $B$                 & 452  & 14414 / 0.01 & 7813 / 0.01   \\
VLT, ESO \citep{nardiello2015}                                    & $B$                 & 452  & 1006 / 0.01 & --       \\
1-m Swope telescope, Las Campanas \citep{narloch2017}             & $B$                 & 452  & -- & 9892 / 0.01  \\
0.9-m CTIO telescope \citep{kaluzny1997}                          & $B$                 & 452  & 4496 / 0.03 & -- \\
2.5-m Du Pont telescope, Las Campanas \citep{kaluzny2010}         & $B$                 & 452  & -- & 7271 / 0.01 \\
2.5-m Du Pont telescope, Las Campanas \citepalias{mandushev1998}  & $B$                 & 452  & -- & fiducial \\
{\it HST}/WFC3 \citep{libralato2022}                              & $F467M$             & 467  & 7805 / 0.01 & --  \\
1.54-m Danish telescope, ESO, La Silla \citepalias{grundahl1999}  & Str\"omgren $b$     & 467  & 5714 / 0.01 & --   \\
0.9-m CTIO telescope \citepalias{twarog2000}                      & Str\"omgren $b$     & 467  & 2476 / 0.01 & -- \\
{\it HST}/ACS \citepalias{simioni2018}                            & $F475W$             & 475  & 4890 / 0.06 & 5627 / 0.07 \\
Pan-STARRS \citep{chambers2016}                                   & $g_\mathrm{PS1}$    & 496  & -- & 4326 / 0.01  \\ 
{\it Gaia} DR3 \citep{riello2021}                                 & $G_\mathrm{BP}$     & 505  & 17312 / 0.02 & 8828 / 0.03   \\
SkyMapper Sky Survey DR3 \citep{onken2019}                        & $g_\mathrm{SMSS}$   & 514  & 7337 / 0.02 & 3332 / 0.02   \\ 
1.54-m Danish telescope, ESO, La Silla \citepalias{grundahl1999}  & Str\"omgren $y$     & 548  & 5713 / 0.01 & --  \\
0.9-m CTIO telescope \citepalias{twarog2000}                      & Str\"omgren $y$     & 548  & 2476 / 0.01 & -- \\
{\it HST}/WFPC2 \citep{piotto2002}                                & $F555W$             & 551  & 5407 / 0.02 & --  \\
Various \citepalias{stetson2019}                                  & $V$                 & 552  & 14432 / 0.01 & 7816 / 0.01  \\
VLT, ESO \citep{nardiello2015}                                    & $V$                 & 552  & 1020 / 0.01 & --  \\
1-m Swope telescope, Las Campanas \citep{narloch2017}             & $V$                 & 552  & -- & 9892 / 0.01  \\
0.9-m CTIO telescope \citep{kaluzny1997}                          & $V$                 & 552  & 4496 / 0.02 & --  \\
2.5-m Du Pont telescope, Las Campanas \citep{kaluzny2010}         & $V$                 & 552  & -- & 7271 / 0.01 \\
0.9-m CTIO telescope \citepalias{twarog2000}                      & $V$                 & 552  & 2476 / 0.01 & -- \\
2.5-m Du Pont telescope, Las Campanas \citepalias{mandushev1998}  & $V$                 & 552  & -- & fiducial \\
Two telescopes \citep{ahumada2021}                                & $V$                 & 552  & 12878 / 0.02 & -- \\
{\it HST}/ACS \citepalias{nardiello2018}                          & $F606W$             & 599  & 12386 / 0.01 & 21417 / 0.01  \\
{\it HST}/ACS \citep{richer2008}                                  & $F606W$             & 599  & 2324 / 0.01 & --  \\
{\it Gaia} DR3 \citep{riello2021}                                 & $G$                 & 604  & 17312 / 0.01 & 8828 / 0.01  \\
SkyMapper Sky Survey DR3 \citep{onken2019}                        & $r_\mathrm{SMSS}$   & 615  & 8581 / 0.02 & 3805 / 0.02    \\
{\it HST}/ACS \citep{libralato2022}                               & $F625W$             & 633  & 12531 / 0.01 & -- \\
Various \citepalias{stetson2019}                                  & $R$                 & 659  & 7730 / 0.01 & 7642 / 0.01    \\
Pan-STARRS \citep{chambers2016}                                   & $i_\mathrm{PS1}$    & 752  & -- & 4326 / 0.01  \\ 
{\it Gaia} DR3 \citep{riello2021}                                 & $G_\mathrm{RP}$     & 770  & 17312 / 0.01 & 8828 / 0.02  \\
SkyMapper Sky Survey DR3 \citep{onken2019}                        & $i_\mathrm{SMSS}$   & 776  & 8387 / 0.02 & 4069 / 0.02    \\
{\it HST}/ACS \citepalias{nardiello2018}                          & $F814W$             & 807  & 12346 / 0.01 & 21417 / 0.01 \\
{\it HST}/ACS \citepalias{simioni2018}                            & $F814W$             & 807  & 4890 / 0.03 & 5627 / 0.03 \\
{\it HST}/ACS \citep{richer2008}                                  & $F814W$             & 807  & 2324 / 0.01 & --  \\
{\it HST}/ACS \citep{libralato2022}                               & $F814W$             & 807  & 12024 / 0.01 & --  \\
Various \citepalias{stetson2019}                                  & $I$                 & 807  & 14423 / 0.01 & 7816 / 0.01   \\
VLT, ESO \citep{nardiello2015}                                    & $I$                 & 807  & 960 / 0.01 & --   \\
2.5-m Du Pont telescope, Las Campanas \citepalias{mandushev1998}  & $I$                 & 807  & -- & fiducial \\
Two telescopes \citep{ahumada2021}                                & $I$                 & 807  & 12878 / 0.02 & -- \\
SkyMapper Sky Survey DR3 \citep{onken2019}                        & $z_\mathrm{SMSS}$   & 913  & 7502 / 0.02 & 3324 / 0.02    \\
2MASS \citep{2mass}                                               & $J_\mathrm{2MASS}$  & 1234 & 4652 / 0.08 & --  \\
VISTA VHS DR5 \citep{vista}                                       & $J_\mathrm{VISTA}$  & 1277 & 17984 / 0.02 & 9292 / 0.02  \\
VISTA VHS DR5 \citep{vista}                                       & $Ks_\mathrm{VISTA}$ & 2148 & 16182 / 0.05 & 6753 / 0.07    \\
{\it WISE}, unWISE \citep{unwise}                                 & $W1$                & 3317 & 1052 / 0.01  & 1116 / 0.01  \\
\hline
\end{tabular}
\]
\end{table*}

Each star has photometry in some but not all filters. In total, 32 and 23 filters are used for NGC\,6397 and NGC\,6809, respectively, spanning a wavelength range between the UV 
and middle IR.
Table~\ref{filters} presents the effective wavelength $\lambda_\mathrm{eff}$ in nm, number of stars and the median photometric precision (after the cleaning of the data sets,
which is described below) for each filter. 
We calculate the median precision from the precision statements by the authors of the data sets.
Then we apply it to evaluate the uncertainties of our results (see appendix A of \citetalias{ngc6205} and Sect.~\ref{results}).

To clean the data sets,
 we generally follow the recommendations of their authors to select single star-like objects with reliable photometry. 
Typically, stars with a photometric uncertainty $<0.12$~mag are selected, while for some data sets we apply a higher or lower cut level between 0.08 and 0.2 mag.
For the {\it HST} WFC3 and ACS photometry, we use stars with $|{\tt sharp}|<0.15$, membership probability $>0.9$ or $-1$, and quality fit $>0.9$.
For the \citetalias{stetson2019} data sets, we use stars with \verb"DAOPHOT" parameters $\chi<3$ and $|{\tt sharp}|<0.3$.
For the SMSS DR3 data sets, we select star-like objects (i.e. with ClassStar$>0.5$) and with flags $<8$.
For the data set of \citetalias{grundahl1999}, we select stars with $\chi<3$ and $|{\tt sharp}|<0.3$.
For the data set of \citet{nardiello2015}, we select stars with the quality of point spread function fit parameter $<0.5$.
For the data set of \citet{kaluzny1997}, we select stars with all quality flags 0.
For the \citet{narloch2017} data set, we use stars with a cluster member probability higher than 0.5.

\begin{figure}
\includegraphics{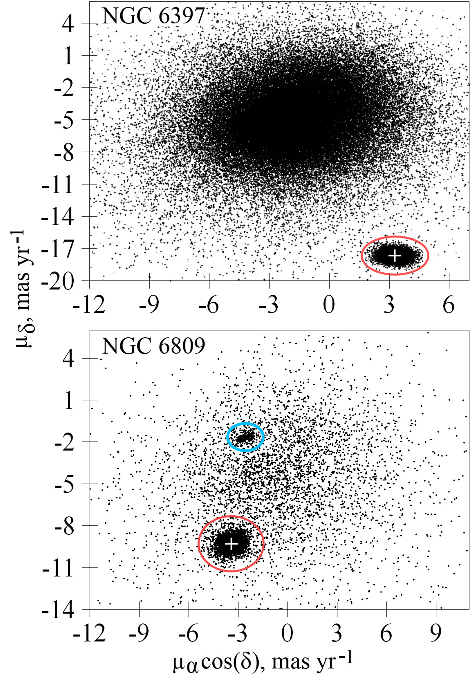}
\caption{The distribution of the {\it Gaia} DR3 data sets, selected within the truncation radii of the clusters, over the PM components (mas\,yr$^{-1}$), 
after the remaining cleaning of the sample. 
To make the figure clearer, we only show stars with precise PM components ($<1.5$ mas\,yr$^{-1}$) and photometry in all filters ($<0.07$ and $<0.1$ mag for NGC\,6397 and 
NGC\,6809, respectively). 
The weighted mean PM and the selection area for the clusters are shown by the white crosses and red circles, respectively.
The blue circle shows a concentration of the Sagittarius dwarf galaxy members behind NGC\,6809.
}
\label{mu}
\end{figure}

\begin{figure}
\includegraphics{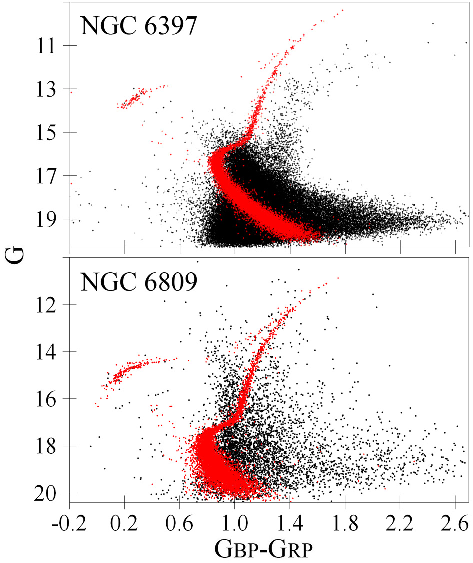}
\caption{$G_\mathrm{BP}-G_\mathrm{RP}$ versus $G$ CMDs for the stars included (red symbols) and excluded (black symbols) by their parallaxes and PMs after the remaining cleaning 
of the {\it Gaia} DR3 samples. For clearer figure, only stars with precise PM components ($<1.5$ mas\,yr$^{-1}$) and photometry in all filters ($<0.07$ and $<0.1$ mag for NGC\,6397 
and NGC\,6809, respectively) are shown. A significant background of the clusters is the MS stars (at $G_\mathrm{BP}-G_\mathrm{RP}\approx1$ and $G<16$ mag) and giants 
(at $G_\mathrm{BP}-G_\mathrm{RP}\approx1.4$ and $G<16$ mag) of Sagittarius dwarf galaxy.  
}
\label{initcmd}
\end{figure}

\begin{figure}
\includegraphics{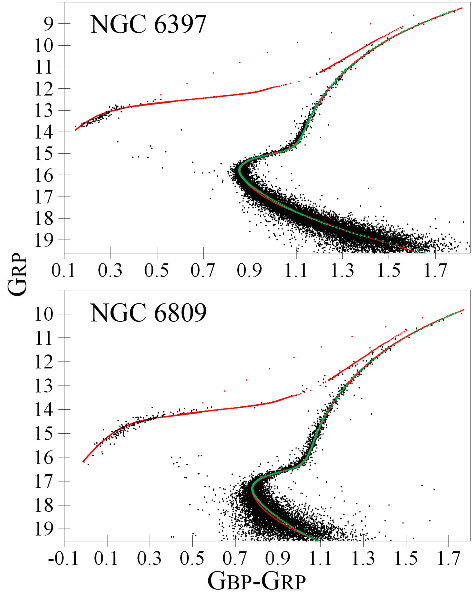}
\caption{$G_\mathrm{BP}-G_\mathrm{RP}$ versus $G_\mathrm{RP}$ CMDs for the {\it Gaia} DR3 clusters members after correction for differential reddening.
The isochrones from BaSTI (red) and DSED (green) for $Y=0.25$ are calculated with the best-fitting parameters from Table~\ref{cmds}.
}
\label{gaia}
\end{figure}

\subsection{{\it Gaia} DR3 cluster members}
\label{edr3}

Similar to \citetalias{ngc6362}, accurate {\it Gaia} DR3 parallaxes and PMs are used to select cluster members and derive systemic parallaxes and PMs.
The distribution of the {\it Gaia} DR3 data sets, selected within the truncation radii of the clusters, over the PM components is presented in Fig.~\ref{mu}.
The cluster members are those inside the red circles. 
It is seen that members can be separated from the fore- and background stars. 

We now briefly describe the selection of the members.
As seen in Table~\ref{properties}, \citet{moreno2014} and \citet{bica2019} provide different estimates for the tidal radii of these GCs. 
Therefore, first, we consider initial {\it Gaia} DR3 samples within initial radii which exceed any previous estimate.
The cluster centre coordinates are taken from \citet{goldsbury2010}.

Second, we find empirical truncation radii of 41 and 18 arcmin for NGC\,6397 and NGC\,6809, respectively, as the radii where the star count surface density drops to 
the Galactic background.
All the data sets, except for the \citetalias{mandushev1998} data set with fiducial sequences only, are truncated at these radii to reduce contamination from non-members.

Third, we leave only stars with PMs;
 \verb|duplicated_source|$=0$ (\verb|Dup=0|), i.e. sources without multiple source identifiers; \verb|astrometric_excess_noise|$<1$ ($\epsilon i<1$);
a renormalized unit weight error not exceeding $1.4$ (\verb|RUWE|$<1.4$); available data in all three {\it Gaia} filters with a precision $<0.12$ mag;
and a corrected excess factor \verb"phot_bp_rp_excess_factor" (i.e. \verb"E(BP/RP)Corr") between $-0.14$ and $0.14$ \citep{riello2021}.
Note that this cleaning removes almost all stars of the {\it Gaia} DR3 data sets within a central arcminute of both the cluster fields.

Fourth, foreground and background stars are rejected as those with an inappropriate parallax (see \citetalias{ngc6362}).

Fifth, we begin with the initial systemic PM components $\overline{\mu_{\alpha}\cos(\delta)}$ and $\overline{\mu_{\delta}}$ from \citet[][hereafter VB21]{vasiliev2021},
calculate the standard deviations $\sigma_{\mu_{\alpha}\cos(\delta)}$ and $\sigma_{\mu_{\delta}}$ 
of the PM components for the cluster members, cut off the sample at $3\sigma$, and recalculate the weighted mean systemic PM components.
We repeat this procedure iteratively until it stops losing stars in the $3\sigma$ cut.
Faint cluster members with less certain PMs make a negligible contribution to the weighted mean systemic PMs.

The final empirical standard deviations $\sigma_{\mu_{\alpha}\cos(\delta)}=0.38$ and $\sigma_{\mu_{\delta}}=0.33$ mas\,yr$^{-1}$ for NGC\,6397 and NGC\,6809, respectively 
(averaged for the PM components) are reasonable, but significantly higher than the mean stated PM uncertainties (0.15 and 0.21 mas\,yr$^{-1}$, respectively),
which may mean an underestimation of the latters.

The CMDs of the {\it Gaia} DR3 stars from Fig.~\ref{mu} are shown in Fig.~\ref{initcmd}.
It is seen that Sagittarius dwarf galaxy is a major background contaminant for both the clusters. For NGC\,6809 it is clearly seen in Fig.~\ref{mu}.
Fig.~\ref{initcmd} shows that the galaxy's members also dominate among bright non-members ($G<16$ mag) of NGC\,6397: the RGB of the galaxy is seen as a bulk of stars 
several magnitudes fainter than the RGB of NGC\,6397.

\begin{table*}
\def\baselinestretch{1}\normalsize\normalsize
\caption{The results of our isochrone fitting for two models and some key CMDs for both the clusters. The colour is the abscissa and the magnitude in the redder filter is the ordinate
in all the CMDs, except the \citetalias{twarog2000} CMDs where $u-v$, $v-b$ or $b-y$ are the abscissas and $V$ is the ordinate. 
Each derived reddening is followed by its empirical systematic uncertainty and corresponding $E(B-V)$, given in parentheses and calculated using extinction coefficients from 
\citet{casagrande2014,casagrande2018a,casagrande2018b} or \citetalias{ccm89} with $R_\mathrm{V}=3.1$.
[Fe/H] is given only for CMDs, which allow its calculation as a fitting parameter. 
The complete table is available online.
}
\label{cmds}
\[
\begin{tabular}{lcccc}
\hline
\noalign{\smallskip}
   & \multicolumn{2}{c}{NGC\,6397} & \multicolumn{2}{c}{NGC\,6809} \\
\noalign{\smallskip}
Quantity                     & DSED                  &  BaSTI                & DSED                  &  BaSTI  \\
\hline
\multicolumn{5}{c}{\citetalias{nardiello2018}} \\
$E(F606W-F814W)$, mag                 & $0.199\pm0.04$ (0.19) & $0.184\pm0.04$ (0.17) & $0.128\pm0.03$ (0.12) & $0.113\pm0.03$ (0.11) \\
age, Gyr	                      & 13.5                  & 13.5                  & 13.0                  & 13.0                  \\
distance, kpc                         & 2.49                  & 2.46                  & 5.22                  & 5.18                  \\
$$ [Fe/H]                             & $-2.0$                & $-1.9$                & $-1.8$                & $-1.7$                \\
\noalign{\smallskip}
\multicolumn{5}{c}{{\it Gaia} DR3} \\ 
$E(G_\mathrm{BP}-G_\mathrm{RP})$, mag & $0.320\pm0.03$ (0.21) & $0.288\pm0.03$ (0.19) & $0.232\pm0.03$ (0.15) & $0.198\pm0.03$ (0.13) \\
age, Gyr                              &   12.5                & 13.0                  & 12.5                  & 13.0                  \\
distance, kpc                         &   2.39                & 2.34                  & 5.04                  & 5.00                  \\
$$ [Fe/H]                             & $-1.8$                & $-1.8$                & $-1.7$                & $-1.7$                \\
\noalign{\smallskip}
\multicolumn{5}{c}{\citetalias{stetson2019}} \\ 
$E(B-V)$, mag                         & $0.162\pm0.03$ (0.18) & $0.157\pm0.03$ (0.18) & $0.107\pm0.03$ (0.12) & $0.098\pm0.03$ (0.11) \\
age, Gyr                              & 13.0                  & 13.0                  & 13.5                  & 13.5                  \\
distance, kpc                         & 2.43                  & 2.44                  & 5.12                  & 5.18                  \\
$$ [Fe/H]                             & $-1.8$                & $-1.8$                & $-1.8$                & $-1.8$                \\
\ldots & \ldots & \ldots & \ldots & \ldots \\
\hline
\end{tabular}
\]
\end{table*}

\begin{table}
\def\baselinestretch{1}\normalsize\small
\caption[]{The cluster systemic PMs (mas\,yr$^{-1}$). The statistic uncertainties are given for the PMs from this study and from \citet{vitral2021}, while the total 
(statistic plus systematic) uncertainty is given for the PMs from \citetalias{vasiliev2021}.
The latter is adopted by us as the most realistic estimate of the PM uncertainties. 
}
\label{systemic}
\[
\begin{tabular}{llcc}
\hline
\noalign{\smallskip}
Cluster & Source & $\mu_{\alpha}\cos(\delta)$ & $\mu_{\delta}$ \\
\hline
\noalign{\smallskip}
          & This study                & $3.260\pm0.010$ & $-17.660\pm0.010$ \\
NGC\,6397 & \citetalias{vasiliev2021} & $3.260\pm0.023$ & $-17.665\pm0.022$ \\
          & \citet{vitral2021}        & $3.256\pm0.003$ & $-17.654\pm0.003$ \\
\noalign{\smallskip}
          & This study                & $-3.430\pm0.010$ & $-9.310\pm0.010$ \\
NGC\,6809 & \citetalias{vasiliev2021} & $-3.431\pm0.025$ & $-9.311\pm0.024$ \\
          & \citet{vitral2021}        & $-3.431\pm0.003$ & $-9.315\pm0.003$ \\
\hline
\end{tabular}
\]
\end{table}

Fig.~\ref{gaia} presents the final $G_\mathrm{BP}-G_\mathrm{RP}$ versus $G_\mathrm{RP}$ CMDs for the {\it Gaia} DR3 clusters members after correction for differential reddening 
described in Sect.~\ref{difred}.

Our final weighted mean systemic PMs are presented in Table~\ref{systemic} in comparison to those from \citetalias{vasiliev2021} and \citet{vitral2021}.
Being obtained from {\it Gaia} DR3 by different approaches, these estimates are, nevertheless,
 consistent within $\pm0.01$ mas\,yr$^{-1}$, i.e. well beneath the {\it Gaia} DR3 PM systematic errors (about 0.02 mas\,yr$^{-1}$), which are estimated by \citetalias{vasiliev2021}.
Since only the statistic uncertainties are evaluated for ours and \citet{vitral2021}'s estimates, we adopt the dominating systematic uncertainties as the final ones of our PMs.

\begin{table}
\def\baselinestretch{1}\normalsize\normalsize
\caption[]{Various parallax estimates (mas) for NGC\,6397 and NGC\,6809 with their total (statistic and systematic) uncertainties.
}
\label{parallax}
\[
\begin{tabular}{lcc}
\hline
\noalign{\smallskip}
 Parallax            &  NGC\,6397  &  NGC\,6809 \\
\hline
\noalign{\smallskip}
\citetalias{vasiliev2021}, {\it Gaia} astrometry  & $0.414\pm0.010$ & $0.206\pm0.010$  \\
This study, {\it Gaia} astrometry                 & $0.416\pm0.010$ & $0.203\pm0.010$  \\
This study, isochrone fitting                     & $0.408\pm0.014$ & $0.191\pm0.007$  \\ 
\hline
\end{tabular}
\]
\end{table}

Similarly, we adopt the total uncertainty of {\it Gaia} DR3 parallaxes, found by \citetalias{vasiliev2021} as 0.01 mas, for our median parallaxes of cluster members.
We correct them for the parallax zero-point following \citet{lindegren2021} and present them in Table~\ref{parallax} for comparison with other estimates in Sect.~\ref{distance}.
Table~\ref{gaiaedr3} contains the final lists of the {\it Gaia} DR3 cluster members.

\begin{table}
\def\baselinestretch{1}\normalsize\normalsize
\caption[]{The list of the {\it Gaia} DR3 members of NGC\,6397 and NGC\,6809. The complete table is available online.
}
\label{gaiaedr3}
\[
\begin{tabular}{cc}
\hline
\noalign{\smallskip}
NGC\,6397 & NGC\,6809 \\
\hline
\noalign{\smallskip}
5921306965291992704 & 6751230058605062400 \\
5921306999650258688 & 6751240954939935872 \\
5921307553705108864 & 6751241053721334144 \\
5921307789925746304 & 6751241058019152384 \\
5921308137820609792 & 6751241500397868928 \\
\ldots & \ldots \\
\hline
\end{tabular}
\]
\end{table}

\begin{figure}
\includegraphics{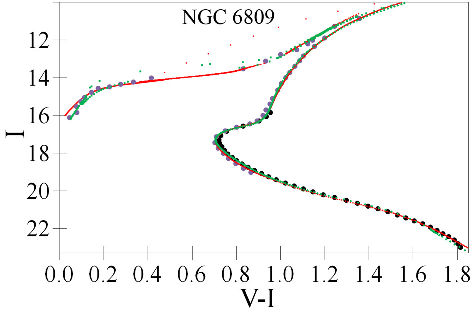}
\caption{$V-I$ versus $I$ CMD with the fiducial sequences from the \citet{mandushev1996} (black circles) and \citet{mandushev1998} (purple circles) observational run for NGC\,6809.
The isochrones from BaSTI (red) and DSED (green diamonds) for $Y=0.25$ are calculated with the best-fitting parameters from Table~\ref{cmds}.
Appropriate DSED predictions for the HB and AGB are shown by the green squares for illustration purposes.
}
\label{mandushev_vi}
\end{figure}

\begin{figure}
\includegraphics{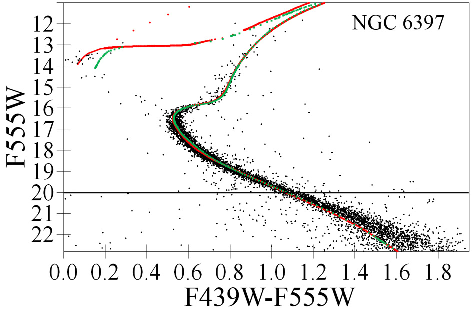}
\caption{{\it HST}/WFPC2 $F439W-F555W$ versus $F555W$ CMD from the \citet{piotto2002} data set for NGC\,6397.
The isochrones from BaSTI (red) and DSED (green diamonds) for $Y=0.25$ are calculated with the best-fitting parameters from Table~\ref{cmds}.
Appropriate DSED predictions for the HB and AGB are shown by the green squares for illustration purposes.
The black horizontal line shows the cut of the faintest MS stars without a reliable isochrone fitting.
}
\label{piotto}
\end{figure}

\begin{figure}
\includegraphics{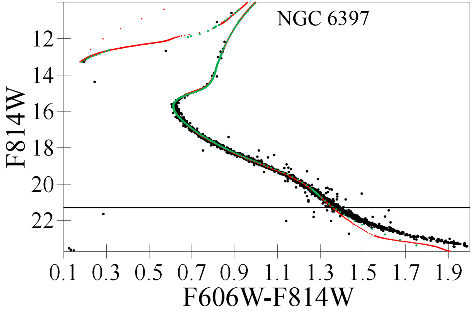}
\caption{The same as Fig.~\ref{piotto} but for the {\it HST}/ACS $F606W-F814W$ colour from the \citet{richer2008} data set for NGC\,6397.
The black horizontal line shows the cut of the faintest MS stars without a reliable isochrone fitting.
}
\label{richer}
\end{figure}

\begin{figure}
\includegraphics{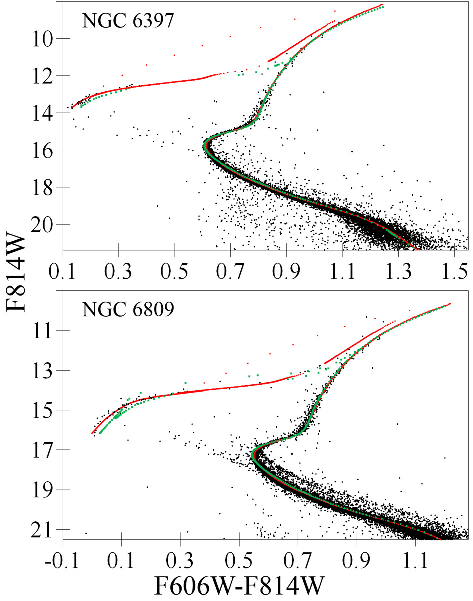}
\caption{{\it HST}/ACS $F606W-F814W$ versus $F814W$ CMD from the \citepalias{nardiello2018} data set.
The isochrones from BaSTI (red) and DSED (green diamonds) for $Y=0.25$ are calculated with the best-fitting parameters from Table~\ref{cmds}.
Appropriate DSED predictions for the HB and AGB are shown by the green squares for illustration purposes.
}
\label{hst}
\end{figure}

\begin{figure}
\includegraphics{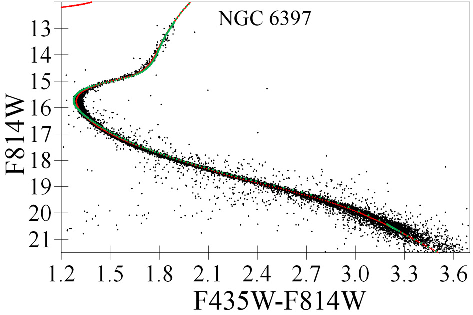}
\caption{The same as Fig.~\ref{gaia} but for the {\it HST}/ACS $F435W-F814W$ colour for NGC\,6397 from \citet{libralato2022}.
}
\label{libralato2022_435_814}
\end{figure}

\begin{figure}
\includegraphics{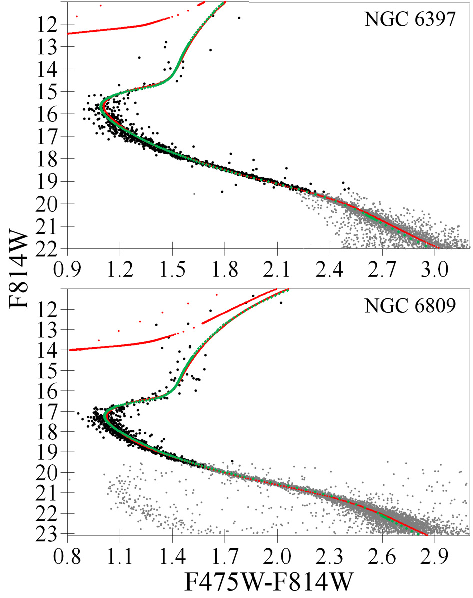}
\caption{The same as Fig.~\ref{gaia} but for the {\it HST}/ACS $F475W-F814W$ versus $F814W$ CMDs with the {\it Gaia} DR3 cluster members from the \citetalias{simioni2018} data sets 
(black) and remaining faint stars from the \citetalias{simioni2018} data sets (grey).
}
\label{simioni}
\end{figure}

\begin{figure*}
\includegraphics{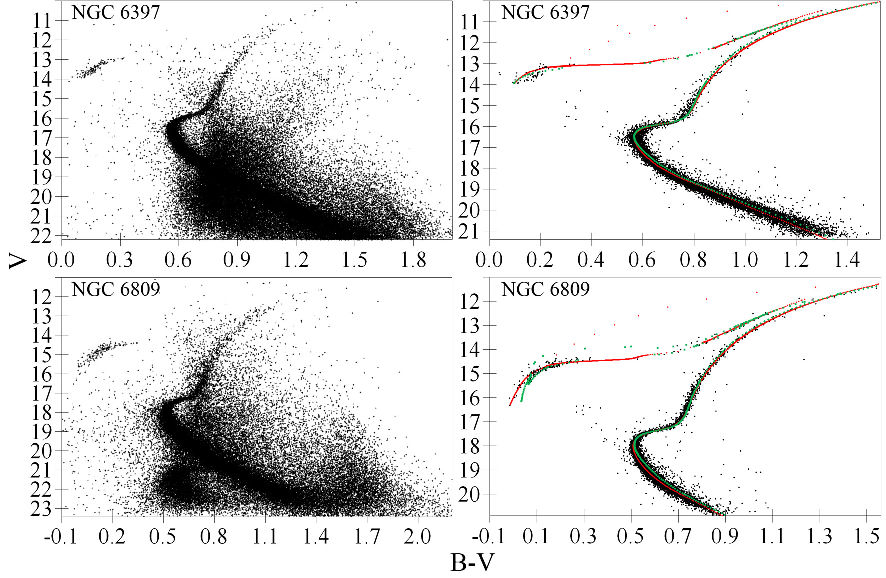}
\caption{$B-V$ versus $V$ CMDs with the initial \citetalias{stetson2019} data sets (left) and the {\it Gaia} DR3 cluster members from the \citetalias{stetson2019} data sets (right).
The isochrones from BaSTI (red) and DSED (green diamonds) for $Y=0.25$ are calculated with the best-fitting parameters from Table~\ref{cmds}.
Appropriate DSED predictions for the HB and AGB are shown by the green squares for illustration purposes.
}
\label{stetson}
\end{figure*}

\subsection{Cluster members in other data sets}
\label{members}

Almost all authors of the data sets under consideration made an effort to select cluster members. This cleaning is acceptable for the 
\citetalias{mandushev1998} and \citet{piotto2002} data sets. The original star-by-star data for the \citetalias{mandushev1998} data set are not available
(we use its fiducial sequences).
Anyway, we cannot cross-identify both the data sets with {\it Gaia} to improve cluster member selection. 
This may introduce some additional systematic errors into corresponding results.
The level of these errors is estimated from the comparison of our results for various data sets in Sect.~\ref{results}.
The CMDs for these data sets are presented in Figs~\ref{mandushev_vi} and \ref{piotto}, respectively. 
Two observational runs of \citet{mandushev1996} and \citet{mandushev1998}, shown by different colours, indicate an insignificant systematic difference of 0.019 mag between their 
TO colours.
We cannot fit the \citet{piotto2002} faintest MS stars by any reliable BaSTI or DSED isochrone and, hence, ignore these stars.

\citet{richer2008}, \citetalias{nardiello2018}, and \citet{libralato2022} have cleaned their {\it HST} data sets from non-members by use of dedicated {\it HST} PMs:
their CMDs are presented in Figs~\ref{richer}, \ref{hst}, and \ref{libralato2022_435_814}, respectively.
Although imperfect, their membership selection cannot be significantly improved through the use of the {\it Gaia} data, since these data sets cover only small fields with 
few {\it Gaia} stars.

Similar to the \citet{piotto2002} faintest MS stars, we cannot fit those of the \citet{richer2008} by any reliable BaSTI or DSED isochrone and, hence, ignore the faintest MS stars.
For the \citet{richer2008} data set this magnitude limit is about $F814W<21.3$. 
The other {\it HST}/ACS data sets of \citetalias{nardiello2018} and \citet{libralato2022} are cut at the same magnitude due to observational limit. 
However, the {\it HST}/ACS data sets of \citetalias{simioni2018}, whose CMDs are presented in Fig.~\ref{simioni}, allows a precise isochrone-to-data fitting by BaSTI and DSED down to a
fainter $F814W\approx23$ mag. This may mean a systematic difference between the \citet{richer2008} and \citetalias{simioni2018} faintest MS stars probably due to systematic errors 
in the former (however, see discussion in \citealt{dicriscienzo2010}).
Anyway, the faint MS slope within about $19<F814W<21.3$ allows us to derive [Fe/H] estimates for the 
\citet{richer2008}, \citetalias{nardiello2018}, \citet{libralato2022}, and \citetalias{simioni2018} data sets (see Table~\ref{cmds}).

The {\it Gaia} DR3 cluster members are among only bright stars of the data sets of \citetalias{simioni2018}, while faint stars of these data sets draw rather clear CMDs.
Therefore, we decide to derive cluster parameters from combined CMDs: we use only {\it Gaia} DR3 cluster members among bright \citetalias{simioni2018} stars (about $F814W<19.5$) 
together with all faint \citetalias{simioni2018} stars, as shown in Fig.~\ref{simioni}.

Cluster members in the \citet{ahumada2021} data set are reliably found by its authors using the method of \citet{bustos2019} after cross-identification with {\it Gaia} DR2.

The remaining data sets are cross-identified with those of {\it Gaia} DR3 to reveal cluster members.
The improvement is seen from a typical Fig.~\ref{stetson}, where contaminated CMDs of NGC\,6397 and NGC\,6809 for the \citetalias{stetson2019} data sets are compared with the same 
CMDs for only {\it Gaia} DR3 members of these data sets. However, this improvement comes at the expense of a few faint magnitudes lost.

The {\it Gaia} membership identification is especially important for NGC\,6397 and NGC\,6809 in order to overcome a bias due to a non-uniform distribution of the Sagittarius 
dwarf galaxy stars over their CMDs, as seen in Fig.~\ref{stetson}.

\begin{figure}
\includegraphics{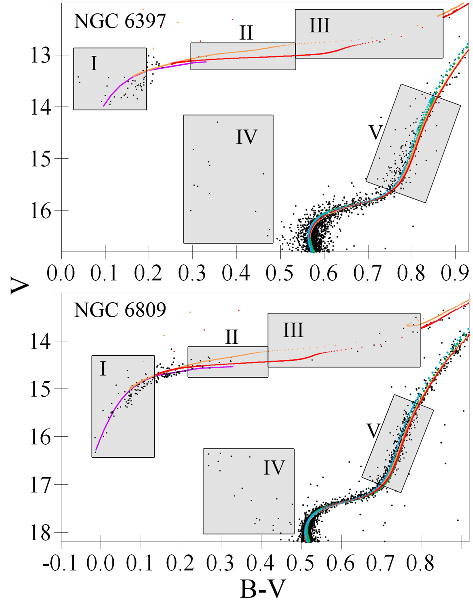}
\caption{A central part of the $B-V$ versus $V$ CMDs with the {\it Gaia} DR3 cluster members from the \citetalias{stetson2019} data sets (i.e. the data from Fig.~\ref{stetson}).
The isochrones from BaSTI (red) and DSED (green) for $Y=0.25$, the BaSTI HB for $Y=0.25$ (magenta), and isochrones from BaSTI (orange) and DSED (blue) for $Y=0.267$ are 
calculated with the best-fitting parameters from Table~\ref{cmds}.
The grey areas are the CMD domains of the (I) blue HB, (II) blue AGB, (III) RR~Lyrae, (IV) blue stragglers, (V) and faint RGB, which are discussed in the text.
}
\label{stetson_details}
\end{figure}

\subsection{Isochrone-to-data fitting}
\label{fitting}

Owing to the accurate selection of the cluster members, our CMDs are more well defined than typical CMDs in the pre-{\it HST} and pre-{\it Gaia} era.
Therefore, we can fit isochrones directly to a bulk of cluster members, without needing to calculate a fiducial sequence.
In this case, the best solution corresponds to a minimal sum of the residuals between isochrone's and data set points.
We select the best isochrone among those calculated for the parameter grid mentioned in Sect.~\ref{iso}.

We have to exclude three CMD domains from the direct fitting: the blue HB, RR~Lyrae variables, and blue stragglers, marked I, III, and IV in Fig.~\ref{stetson_details}, 
respectively.
The blue HB, i.e. the area bluer than the turn of the observed HB downward, is excluded, since even its best prediction deviates from the observations in typical CMD 
when its other domains are fitted well, as seen in Fig.~\ref{stetson_details}.
We fit the HB stars between the areas I and II by the BaSTI HB models with $Y=0.25$, while the stars in the area II (blue AGB) are better fitted by the AGB isochrones with 
higher $Y=0.267$, as noted in Sect.~\ref{iso}.
Another CMD domain fitted with higher $Y=0.267$ is the faint RGB, marked V in Fig.~\ref{stetson_details} and mentioned in Sect.~\ref{iso}.
All the remaining stars are fitted by isochrones with $Y=0.25$.

To balance the contributions of different CMD domains, we assign a weight to each data point. 
The weight is inversely proportional to the number of stars of a given magnitude for a given data set, i.e. it reflects the luminosity function of a given data set.

Since we fit a zigzag pattern of an isochrone to a zigzag pattern of the bulk of stars, different parts of them are more sensitive to different parameters.
Namely, reddening and distance correlate with the overall shift of the pattern along the abscissa (i.e. colour) and ordinate (i.e. magnitude), respectively.
Therefore, nearly vertical and nearly horizontal parts of the pattern are more sensitive to the determination of reddening and distance, respectively.
Similarly, [Fe/H] is more sensitive to the slopes of the RGB and faint MS.
Finally, age correlates with the length of the SGB, as well as with the HB--SGB and SGB--MS magnitude differences, although different definitions of each of these quantities 
are possible (e.g. the SGB--MS magnitude difference can be defined as the one between the middle of the SGB and the MS of the same colour).

We have checked that the results of the isochrone-to-data fitting obtained by two methods, with and without fiducial sequences, almost coincide.
Namely, the derived [Fe/H], ages, distances, and reddenings [converted into $E(B-V)$] agree within 0.1 dex, 0.5 Gyr, 80 pc, and 0.01 mag, respectively.  
The results obtained without fiducial sequences are presented in Table~\ref{cmds}.

\begin{figure*}
\includegraphics{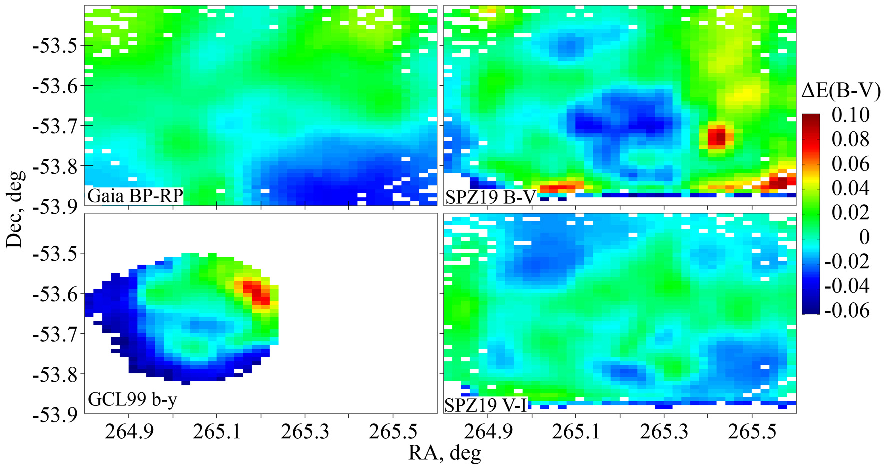}
\caption{The differential reddening maps for the same NGC\,6397 field obtained by use of four CMDs.
}
\label{dr_ngc6397}
\end{figure*}

\begin{figure}
\includegraphics{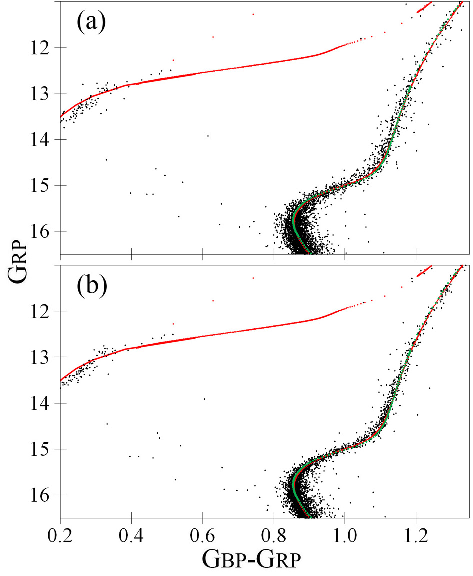}
\caption{A central part of the {\it Gaia} DR3 $G_\mathrm{BP}-G_\mathrm{RP}$ versus $G_\mathrm{RP}$ CMD for NGC\,6397 from Fig.~\ref{gaia} 
(a) before and (b) after differential reddening correction.
}
\label{ngc6397gaia_beforeafter}
\end{figure}

\subsection{Differential reddening}
\label{difred}

Differential reddening (DR) across the fields of both clusters is taken into account following the method of \citetalias{bonatto2013}.
Briefly, the cluster field is divided into a cell grid, with the angular resolution being higher in regions containing more stars. 
Then, the stellar-density Hess diagram (including photometric errors) of each cell is matched to the average (whole field) diagram by applying shifts along the reddening vector 
that are subsequently converted into DR in the cell. By design, the method assumes that differences in the CMDs can be accounted for entirely by DR. 
However, other variations of CMD over cluster field, such as photometry zero-point variations, point-spread function variations, telescope focus change, distortion, 
telescope breathing, stellar population variations, and other reasons, discussed by \citet{anderson2008}, are difficult to separate from DR.

It appears that only data sets with at least 3000 stars provide sufficient coverage of the cluster fields and, hence, draw rather precise DR maps,
i.e. the data sets of
\citetalias{nardiello2018},
\citetalias{grundahl1999},
{\it Gaia} DR3,
\citetalias{stetson2019},
SMSS,
\citet{libralato2022},
\citet{narloch2017},
\citet{kaluzny1997},
\citet{kaluzny2010},
PS1,
2MASS,
VISTA,
and all their cross-identifications,
with the exception of the NGC\,6397 data sets of \citet{piotto2002} and \citet{ahumada2021} with insufficient information about stellar coordinates.

Four examples of DR maps for NGC\,6397 are shown in Fig.~\ref{dr_ngc6397}. All DR maps of both the clusters show that:
\begin{itemize}
\item the DR maps have little to do with each other as for different data sets, as for different CMDs/colours of the same data set;
\item DR variations are mostly small: within $\Delta E(B-V)=\pm0.04$ mag after conversion by use of any reliable extinction law;
\item some CMDs/colours show strong gradients over the fields (e.g. for \citetalias{grundahl1999} $b-y$ in Fig.~\ref{dr_ngc6397}) or
sharp extremes of DR in a small area (e.g. red peaks for \citetalias{stetson2019} $B-V$ in Fig.~\ref{dr_ngc6397}).
\end{itemize}
The peaks in the \citetalias{stetson2019} DR maps can be explained as a manifestation of initial observational data sets (\citetalias{stetson2019} data set combines them)
covering small parts of the field and having significant systematics over the field.
Given that DR is not large across the fields of NGC\,6397 and NGC\,6809, these findings show that the effects mentioned above are more important than DR itself in the CMDs of 
both the clusters.
Anyway, our correction of the data sets for DR reduces the scatter of their stars around their ridge lines or best-fitting isochrones in CMDs, e.g. in Fig.~\ref{ngc6397gaia_beforeafter}.
Note that the mean DR correction for all the CMDs is exactly zero. This leads to a negligible shift of bulk of stars in the CMDs and does not change an average reddening
over the field, which is presented in Table~\ref{cmds}.

Since each data set and CMD/colour draw its own DR map, it is not surprising that our DR maps differ from those of \citet{alonso2012}.

\begin{figure}
\includegraphics{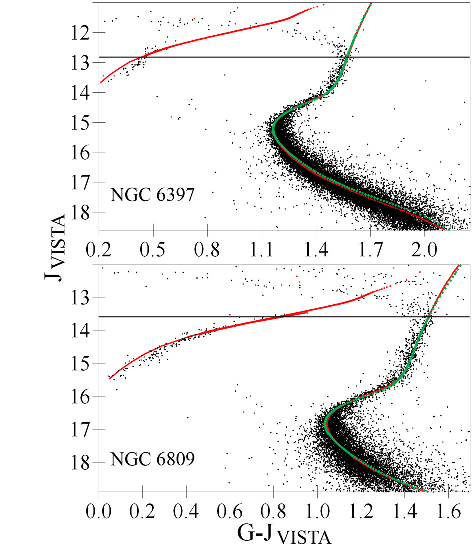}
\caption{The same as Fig.~\ref{gaia} but for the {\it Gaia} -- VISTA $G-J_\mathrm{VISTA}$ colour.
The black horizontal line shows the cut of bright stars with systematically erroneous photometry.
}
\label{gaiavista}
\end{figure}

\begin{figure}
\includegraphics{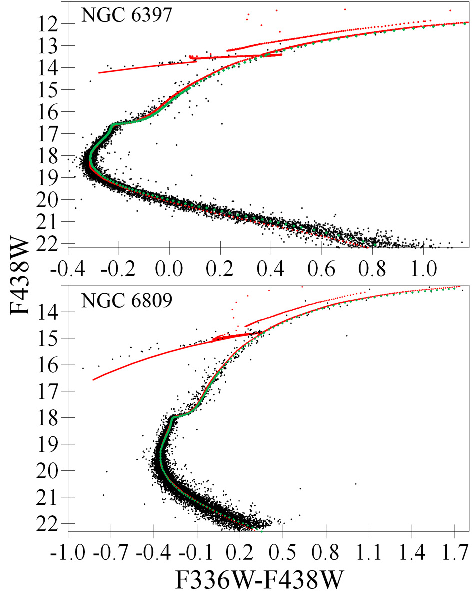}
\caption{The same as Fig.~\ref{gaia} but for {\it HST} $F336W-F438W$ versus $F438$ CMDs \citepalias{nardiello2018}.
}
\label{hst_336_438}
\end{figure}

\begin{figure}
\includegraphics{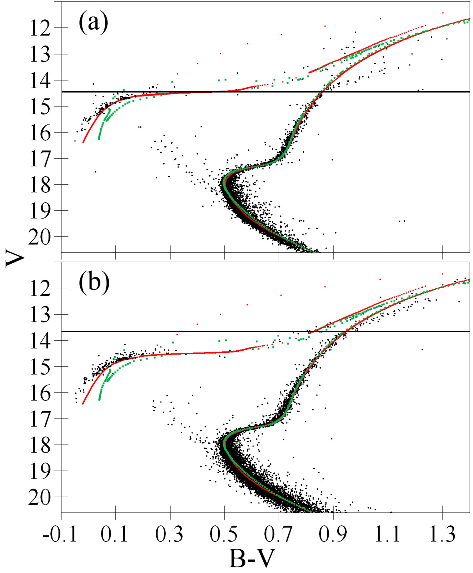}
\caption{The same as Fig.~\ref{piotto} but for the $B-V$ colour from the (a) \citet{kaluzny2010} and (b) \citet{narloch2017} data sets for NGC\,6809.
The black horizontal lines show the cut of bright stars with their biased colour.
}
\label{kaluzny2010_narloch_bv}
\end{figure}

\section{Results}
\label{results}

As in our previous papers, we fit isochrones to a hundred CMDs with different colours. The results for adjacent CMDs appear consistent.
We present some interesting CMDs with isochrone fits in Figs~\ref{gaia}--\ref{stetson} and Figs~\ref{gaiavista}--\ref{kaluzny2010_narloch_bv}. 
Other CMDs are presented online or can be provided on request.

We present the obtained [Fe/H], distances, reddenings, and ages for the most important CMDs in Table~\ref{cmds}.
For comparison, we convert the obtained reddenings into $E(B-V)$, given in parentheses, by use of extinction coefficients from \citet{casagrande2014,casagrande2018a,casagrande2018b} or 
\citet[][hereafter CCM89]{ccm89} with $R_\mathrm{V}=3.1$.\footnote{Extinction-to-reddening ratio $R_\mathrm{V}\equiv A_\mathrm{V}/E(B-V)=3.1$ is defined for early type MS stars, 
while the observed ratio $A_\mathrm{V}/E(B-V)$ depends on intrinsic spectral energy distribution of stars under consideration \citep{casagrande2014}.
For rather cool and metal-poor stars of the GCs under consideration the observed reddening is calculated as $E(B-V)=A_\mathrm{V}/3.48$,
while the extinction coefficients are calculated for the median effective temperature 6400 K of the cluster members.}

Table~\ref{cmds} provides the empirical systematic uncertainty of obtained reddening after its value. 
This systematic uncertainty is defined as the maximal deviation of the best-fitting isochrone from the bulk of the stars along the reddening vector (i.e. nearly along the colour).
The systematic uncertainty never drops below 0.03 mag and it is usually larger than the predicted statistic uncertainty.
The latter is described in the balance of uncertainties (see appendix A of \citetalias{ngc6205}).
The largest values in such pairs of the systematic and statistic uncertainties are shown by the extinction error bars in Figs~\ref{ngc6397law} and \ref{ngc6809law} with
resulting empirical extinction laws.

\subsection{Issues}
\label{issues}

The VISTA photometry for the brightest stars is biased, as discussed in \citetalias{ngc6362}. This is seen in Fig.~\ref{gaiavista}. 
However, accurate parameters of the clusters can be obtained by use of the remaining VISTA stars.

In \citetalias{ngc6362}, we discussed drawbacks of UV, UV--optical, optical--IR, and IR--IR CMDs with respect to (w.r.t.) a typical optical CMD 
(with filters within $430<\lambda_\mathrm{eff}<1000$ nm).
Especially, the UV CMDs are highly affected by the multiple population chemical patterns (see \citealt{sbordone2011} and \citealt{cassisi2013} for the first results,
as well as a recent discussion by \citealt{vandenberg2022}).
However, some of these CMDs can give reliable [Fe/H], age, reddening, and distance estimates (see Table~\ref{cmds}).

Fig.~\ref{hst_336_438} shows an example UV CMD with the {\it HST}/WFC3 $F336W-F438W$ colour, where almost all domains are fitted by both the BaSTI and DSED isochrones 
with reasonable residuals.
The maximal colour offset of the best-fitting isochrones from these data is 0.05 mag and it is the same for both the models and both the clusters.
Moreover, the maximal colour offsets are $<0.05$ mag for the UV CMDs with the $u-v$ Str\"omgren colour from the \citetalias{grundahl1999} and \citetalias{twarog2000} data sets, 
as well as for the CMDs with the $U-B$ colour from the \citetalias{stetson2019} data sets. 
For comparison, such offsets were 0.08--0.10 mag for the UV CMD with {\it HST}/WFC3 $F336W-F438W$ colour for NGC\,6362 and NGC\,6723 in \citetalias{ngc6362}.
Thus, it seems that the BaSTI and DSED UV isochrones better fit UV CMDs for low metallicity GCs (NGC\,6397 and NGC\,6809) than for those with a higher metallicity 
(NGC\,6362 and NGC\,6723).
However, Table~\ref{cmds} shows that some distance and age estimates derived from the UV CMDs are unreliable.
Thus, the results from UV, UV--optical, optical--IR, and IR--IR pairs (including those in Table~\ref{cmds}) are not used for our final estimates.

We do not use the \citetalias{stetson2019} photometry in the $R$ filter for our final estimates due to its lower precision than in the $BVI$ filters.

\begin{figure*}
\includegraphics{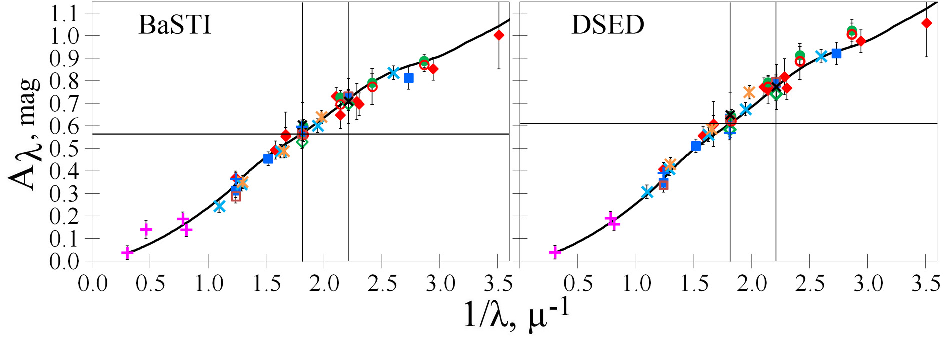}
\caption{The empirical extinction laws for NGC\,6397 from the isochrone fitting drawn by the different models.
The data sets are:
{\it HST} WFC3 and ACS by \citet{richer2008,nardiello2018,simioni2018,libralato2022} -- red diamonds;
{\it Gaia} -- yellow snowflakes;
\citetalias{stetson2019} -- blue squares;
SMSS -- blue inclined crosses;
{\it HST} WFPC2 by \citet{piotto2002} -- black inclined crosses;
\citet{nardiello2015} -- open brown squares;
\citet{kaluzny1997} -- open green diamonds;
\citetalias{grundahl1999} -- green circles;
\citetalias{twarog2000} -- open red circles;
\citet{ahumada2021} -- blue upright crosses;
IR data sets from 2MASS, VISTA, and unWISE -- purple upright crosses.
The vertical lines denote the effective wavelengths of the $B$ and $V$ filters.
The black curve shows the extinction law of \citetalias{ccm89} with $R_\mathrm{V}=3.3$ tied to the obtained $A_\mathrm{V}$, which is shown by the horizontal line.
}
\label{ngc6397law}
\end{figure*}

\begin{figure*}
\includegraphics{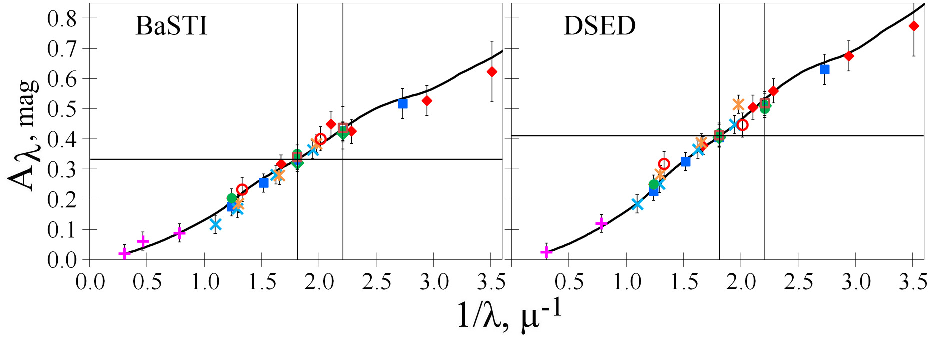}
\caption{The same as Fig.~\ref{ngc6397law} but for NGC\,6809.
The data sets are:
{\it HST} WFC3 and ACS by \citet{nardiello2018,simioni2018} -- red diamonds;
{\it Gaia} -- yellow snowflakes;
\citetalias{stetson2019} -- blue squares;
SMSS -- blue inclined crosses;
\citetalias{mandushev1998} -- green circles
\citet{kaluzny2010} -- open green diamonds;
\citet{narloch2017} -- open brown squares;
PS1 -- open red circles;
IR data sets by VISTA and unWISE -- purple upright crosses.
The lines and curves are the same as in Fig.~\ref{ngc6397law}, but for $R_\mathrm{V}=2.9$.
}
\label{ngc6809law}
\end{figure*}

Similar to \citetalias{ngc6362}, our cross-identification of data sets, which use the same or similar filters, reveals some systematic differences up to 0.04 mag in magnitudes and 
colours for some data sets.
They are common and expected \citepalias{stetson2019}. Our DR corrections reduce these differences and, hence, confirm that they are mostly due to some systematic errors 
of the data sets (see Sect.~\ref{difred}).
We do not take into account the residual systematics after the DR correction, since we find little, if any, influence of these systematics to the derived parameters, 
as seen in Table~\ref{cmds} and in Figs~\ref{ngc6397law} and \ref{ngc6809law}.
In particular, similar to \citetalias{ngc6362} and in contrast to \citetalias{ngc288}, we do not adjust the data sets. 
The adjustment would slightly decrease the scatter of the derived parameters. However, without the adjustment, we better see the real influence of the data set systematics.
Yet, we eliminate the brightest stars of the \citet{kaluzny2010} and \citet{narloch2017} data sets for NGC\,6809 due to their exceptionally large systematic deviation 
from any reasonable isochrone, as seen in Fig.~\ref{kaluzny2010_narloch_bv} in comparison with Fig.~\ref{stetson} for the \citetalias{stetson2019} data set.

\subsection{Metallicity}
\label{metal}

We obtain [Fe/H]$=-1.86$ and $-1.82$ for NGC\,6397 from DSED and BaSTI, respectively, as the average from 16 independent optical CMDs with the well-populated bright RGB or faint MS.
Their mean [Fe/H]$=-1.84\pm0.02\pm0.1$ is adopted as our final [Fe/H] estimate for NGC\,6397 and used for other CMDs.
The statistic uncertainty $\pm0.02$ is calculated as the standard deviation of one estimate divided by the square root of the number of the estimates, 
while the systematic uncertainty is the uncertainty of the iron scale $0.1$ dex, discussed in Sect.~\ref{clusters}, which is larger than the DSED--BaSTI systematic difference 
of $0.04$ dex.
Similarly, using 10 CMDs for NGC\,6809, we obtain [Fe/H]$=-1.82$ (DSED) and $-1.74$ (BaSTI) and their mean [Fe/H]$=-1.78\pm0.02\pm0.1$ as our final [Fe/H] estimate.

Comparing our [Fe/H] estimates with those from spectroscopy, mentioned in Sect.~\ref{clusters}, we conclude that our
 estimates support the higher estimates from \citet{meszaros2020} ([Fe/H]$=-1.89\pm0.09$ and $-1.76\pm0.07$ for NGC\,6397 and NGC\,6809, respectively) 
and \citet{wang2017} ([Fe/H]$=-1.86\pm0.06$ for NGC\,6809), 
but not the lower ones from \citet{carretta2009} ([Fe/H]$=-1.99\pm0.02$ and $-1.93\pm0.02$ for NGC\,6397 and NGC\,6809, respectively)
and \citet{rain2019} ([Fe/H]$=-2.01\pm0.02$ for NGC\,6809).
Accordingly, following the discussion of \citet{meszaros2020}, our estimates support the reference solar abundance 
mixture from \citet{grevesse2007}, but not from \citet{gratton2003}.

\begin{table}
\def\baselinestretch{1}\normalsize\normalsize
\caption[]{The $E(B-V)$ values found through the various models.
The model estimates are average values for all optical CMDs, provided with the standard deviations of the mean values. 
The final values are the mean values of the models with their uncertainties as half the differences between the model estimates.
}
\label{av}
\[
\begin{tabular}{lcc}
\hline
\noalign{\smallskip}
                     & NGC\,6397         & NGC\,6809       \\
\hline
\noalign{\smallskip}
BaSTI                & $0.172\pm0.006$   & $0.112\pm0.004$  \\    
DSED                 & $0.184\pm0.006$   & $0.125\pm0.004$  \\    
\noalign{\smallskip}
Final value          & $0.178\pm0.006$   & $0.118\pm0.006$ \\   
\hline
\end{tabular}
\]
\end{table}

\subsection{Reddening and extinction}
\label{redext}

We verify the agreement of reddening estimates from all CMDs with each other and with an extinction law by combining all derived reddening estimates into empirical extinction laws.
These laws are presented in Figs~\ref{ngc6397law} and \ref{ngc6809law}.

Similar to our previous papers, in order to draw these laws, we cross-identify all possible data sets with the 2MASS, VISTA and unWISE data sets and 
calculate extinctions in all filters from the derived reddenings and IR extinctions. For example,
\begin{equation}
\label{avaw1}
A_\mathrm{V}=(A_\mathrm{V}-A_\mathrm{W1})+A_\mathrm{W1}=E(V-W1)+A_\mathrm{W1},
\end{equation}
where $E(V-W1)$ is obtained from a CMD, while very low extinction $A_\mathrm{W1}$ in the $W1$ filter is slightly upgraded iteratively with upgrade of
extinction law.

Note that some data set pairs cannot be cross-identified. 
The main reasons are a very small common field or common magnitude range of such data sets.
Namely, the \citetalias{nardiello2018} data set for NGC\,6397 contains only rather faint stars in a small field and, hence, has only a few common stars with 2MASS.
Other such pairs: the \citetalias{nardiello2018} data set for NGC\,6809 versus VISTA and unWISE and \citet{kaluzny2010} data set for NGC\,6809 versus unWISE.
The data sets of \citetalias{mandushev1998}, \citet{piotto2002}, \citet{richer2008}, \citetalias{simioni2018}, and \citet{ahumada2021} are not cross-identified with any IR data set.
The extinctions for the \citetalias{mandushev1998}, \citet{piotto2002}, and \citet{ahumada2021} data sets are calculated by adopting the \citetalias{ccm89} extinction law with 
our best-fitting $R_\mathrm{V}$ (described later) for their $I$, {\it HST}/WFPC2 $F555W$, and $I$ filters, respectively.

The common {\it HST}/ACS $F814W$ filter for the \citetalias{nardiello2018}, \citet{richer2008}, \citetalias{simioni2018}, and \citet{libralato2022} data sets allows us to process 
their reddenings together by adopting the same extinction $A_\mathrm{F814W}$.
Extinctions derived for all {\it HST} filters, detectors, and data sets are shown together in Figs~\ref{ngc6397law} and \ref{ngc6809law} by the red diamonds.
These extinction estimates agree with each other following the same smooth extinction laws without outliers.
This is a robust confirmation of the systematic accuracy of our reddening and extinction estimates at the level of a few hundredths of a magnitude.
Thus, we use all the {\it HST} results for its optical filters ($\lambda_\mathrm{eff}>430$ nm) for our final estimates of [Fe/H], age, reddening, and distance.

2MASS versus unWISE and VISTA versus unWISE are cross-identified via common {\it Gaia} DR3 cluster members. 
The VISTA-unWISE CMDs represent a very short wavelength baseline and, hence, provide uncertain age, distance, and [Fe/H]. 
Therefore, fixing these parameters for the VISTA-unWISE pair, we derive only the reddening $E(J_\mathrm{VISTA}-W1)$ as the average difference w.r.t. $G_\mathrm{RP}$ and 
$V$ from \citetalias{stetson2019}: 
$E(J_\mathrm{VISTA}-W1)=[E(G_\mathrm{RP}-W1)-E(G_\mathrm{RP}-J_\mathrm{VISTA})+E(V-W1)-E(V-J_\mathrm{VISTA})]/2$.

The extinctions in Figs~\ref{ngc6397law} and \ref{ngc6809law} show a low scatter of a few hundredths of a magnitude around the \citetalias{ccm89} extinction law with 
best-fitting $R_\mathrm{V}=3.3$ and 2.9 for NGC\,6397 and NGC\,6809, respectively.

We derive our final reddening and extinction estimates for the clusters through the use of all 19 and 14 independent optical CMDs for NGC\,6397 and NGC\,6809, respectively.
Table~\ref{av} presents the final reddening estimates. The DSED estimates are systematically higher than those from BaSTI by about $\Delta E(B-V)=0.01$ mag.
This is related to the systematically lower [Fe/H] of the DSED best-fitting isochrones.

We calculate our final $A_\mathrm{V}$ estimates as the averages of all its direct measurements (6 for NGC\,6397 and 4 for NGC\,6809) using equation (\ref{avaw1}) and its 
counterparts for other IR filters and the Str\"omgren $y$ filter, which is very close to the $V$ one.
We obtain $A_\mathrm{V}=0.59\pm0.01\pm0.02$ and $0.37\pm0.01\pm0.04$ mag (statistical and model-to-model uncertainties) for NGC\,6397 and NGC\,6809, respectively.
Accordingly, the ratio of these $A_\mathrm{V}$ and $E(B-V)$ estimates, 
$R_\mathrm{V}=3.32^{+0.32}_{-0.28}$ and $3.16^{+0.66}_{-0.56}$ (total uncertainty) for NGC\,6397 and NGC\,6809, respectively.

The systematic uncertainty 0.1 dex for [Fe/H] (see Sect.~\ref{metal}) is the dominant contribution to systematic uncertainty of all our reddening and extinction results, 
which is equivalent to $\sigma_{E(B-V)}=0.01$ and $\sigma_{A_\mathrm{V}}=0.03$.

Our $E(B-V)$ estimates agree with those in Table~\ref{properties} for both the clusters, except very high estimate by \citet{planck} for NGC\,6397 and very low estimate by 
\citet{harris} for NGC\,6809. We find no reason for these outliers.
Also, our $E(B-V)$ estimates agree with those from isochrone fitting in Table~\ref{previous}, except very low $E(B-V)$ by \citet{martinazzi2014} for NGC\,6397 and
by \citet{valcin2020} for NGC\,6809.

Reddening estimates calculated through alternative methods are:
\citet{olech1999} found $E(B-V)=0.11\pm0.03$ for RR~Lyrae variables of NGC\,6809;
\citetalias{twarog2000} found $E(B-V)=0.179\pm0.003\pm0.011$ (statistic and systematic uncertainties) from the Str\"omgren photometric data of the TO stars of NGC\,6397;
\citet{hansen2007} found $E(F606W-F814W)=0.20\pm0.03\approx E(B-V)$ from the white dwarf cooling sequence of NGC\,6397;
\citet{pych2001} with a correction from \citet{mcnamara2011} derived $E(B-V)=0.135\pm0.005$ from the data for SX~Phe variables in NGC\,6809.
All these estimates perfectly agree with ours.

\begin{table}
\def\baselinestretch{1}\normalsize\normalsize
\caption[]{Our NGC\,6397 and NGC\,6809 age (Gyr) and distance (kpc) estimates from optical CMDs. All the uncertainties are standard deviations of one measurement.
}
\label{agedist}
\[
\begin{tabular}{lccc}
\hline
\noalign{\smallskip}
                     & DSED           & BaSTI          &  Mean value \\      
\hline
\noalign{\smallskip}
                     &                & NGC\,6397      & \\
\noalign{\smallskip}
Mean distance        & $2.46\pm0.08$  & $2.44\pm0.08$  & $2.45\pm0.08$ \\  
Mean age             & $12.95\pm0.62$ & $12.89\pm0.59$ & $12.92\pm0.60$ \\ 
\noalign{\smallskip}
                     &                & NGC\,6809      & \\
\noalign{\smallskip}
Mean distance        & $5.26\pm0.12$  & $5.22\pm0.10$  & $5.24\pm0.11$  \\
Mean age             & $12.96\pm0.60$ & $12.96\pm0.54$ & $12.96\pm0.56$ \\  
\hline
\end{tabular}
\]
\end{table}


\subsection{Distance and age}
\label{distance}

We average our distance and age estimates from all 19 and 14 optical CMDs for NGC\,6397 and NGC\,6809, respectively.
Table~\ref{agedist} with the final results shows the consistent standard deviations for the models and for the mean values and, hence, 
a good agreement between BaSTI and DSED in their distance and age estimates.
Our final estimates for NGC\,6397 and NGC\,6809, respectively, are as follows:
\begin{enumerate}
\item age is $12.9\pm0.1\pm0.8$ and $13.0\pm0.1\pm0.8$ Gyr (statistic and systematic uncertainties), 
\item distance is $2.45\pm0.02\pm0.06$ and $5.24\pm0.02\pm0.18$ kpc,
\item distance modulus $(m-M)_0=11.95\pm0.01\pm0.05$ and $13.60\pm0.01\pm0.07$ mag,
\item apparent $V$-band distance modulus $(m-M)_\mathrm{V}=12.54\pm0.03\pm0.06$ and $13.97\pm0.03\pm0.08$ mag.
\end{enumerate}

Despite the rather sparse HB populations in NGC\,6397 and NGC\,6809, the statistic uncertainty of their derived distances is lower than the systematic uncertainty. 
The latter can be estimated from the scatter of the previous estimates of the distance moduli or distances presented in the compilation of all GC distance determinations
by \citet{baumgardt2021}.\footnote{This compilation is so comprehensive that our distance estimates do not need a comparison with individual estimates from the literature.}
All recent (since \citealt{dotter2010}) estimates of the distance modulus for NGC\,6397 by isochrone-to-CMD fitting, except outlying 12.12 from \citet{valcin2020}, are within 
11.95--12.05 (including our own 11.95).
Similarly, those for NGC\,6809, except outlying 13.70 from \citet{valcin2020}, are within 13.53--13.67 (including our own 13.60).
Assuming this scatter is due to some systematics, we accept a systematic uncertainty of distance moduli as $\pm0.05$ and $\pm0.07$ for NGC\,6397 and NGC\,6809, respectively, 
which converts to $\pm62$ and $\pm178$ pc distance systematic uncertainty.
Such large systematics may be due to contamination of the HB and SGB of both the clusters by the MS of Sagittarius dwarf galaxy, as seen in Fig.~\ref{stetson}.
Note that our distance estimates agree with the most probable distance estimates of \citet{baumgardt2021} presented in Table~\ref{properties}: 
within 31 and 108 pc, i.e. $0.8\sigma$ and $1.5\sigma$ of their stated statistical uncertainties for NGC\,6397 and NGC\,6809, respectively, and well inside the systematic uncertainties.
Both ours and \citet{baumgardt2021}'s estimates for both the clusters differ considerably from those of \citet{harris} (see Table~\ref{properties}) and, hence, can update them.

We convert our distance estimates into parallaxes to compare them in Table~\ref{parallax} with our parallax estimates from the {\it Gaia} DR3 astrometry (see Sect.~\ref{edr3}) 
and with those from \citetalias{vasiliev2021}.
A good agreement between the parallaxes within the stated uncertainties is seen.
Note that for such nearby GCs, the parallax estimates from the {\it Gaia} DR3 astrometry have nearly the same precision as those from our isochrone fitting,
unlike more distant GCs in our previous studies, whose isochrone fitting parallaxes are more precise.

The systematic uncertainty of age was discussed and estimated in Section 3.1 of \citetalias{ngc6362}.
This should take into account the scatter of the previous age estimates in Table~\ref{previous} and others 
[e.g. NGC\,6397's age of $12.8\pm0.50\pm0.75$ Gyr (statistic and systematic uncertainties) from a population synthesis study of the white dwarf population by \citet{torres2015} and
$13.4\pm0.7\pm1.2$ from the $V$ luminosity at the TO by \citet{brown2018}].
Taking into account this scatter and the discussion of age uncertainty by \citet{vandenberg2018} and \citet{valcin2020},
we assign 0.8 Gyr as the systematic uncertainty of our derived ages.

\begin{figure*}
\includegraphics{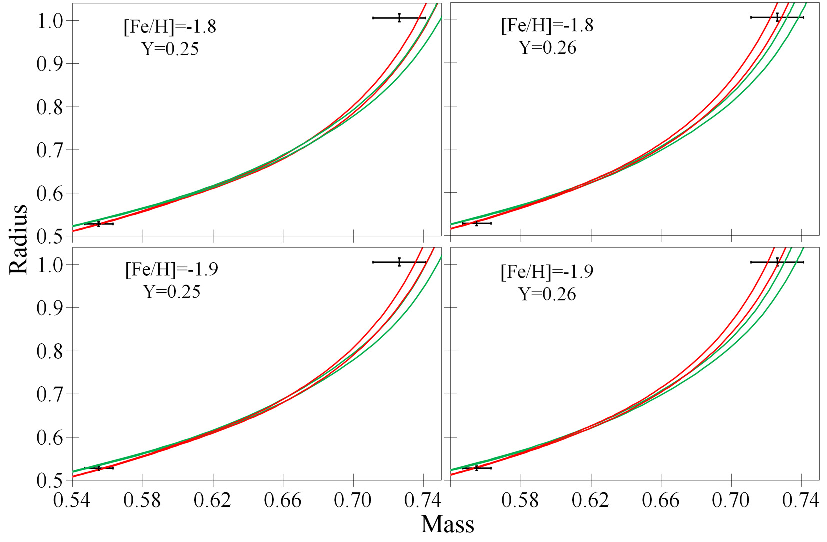}
\caption{The mass--radius relation (in solar mass and radius) for the eclipsing binary V54 of NGC\,6809 (black symbols with error bars) fitted by the BaSTI (red) and DSED (green) 
isochrones for various [Fe/H], $Y$, and age (the left and right isochrone in each model pair is for 13.5 and 13 Gyr, respectively).
}
\label{ngc6809_v54}
\end{figure*}

\subsection{Eclipsing binary}
\label{binary}

NGC\,6809 contains the detached eclipsing binary V54, whose component masses and radii were measured accurately by \citet{kaluzny2014}.
Hence, we can verify the cluster parameters by fitting isochrones to the precise V54 mass--radius relation.

\citet{kaluzny2014} estimated the age, distance modulus, and helium abundance of V54 as 13.3--14.7 Gyr, $(m-M)_V=13.94\pm0.05$ mag, and $Y=0.25$, respectively, 
adopting $E(B-V)=0.115\pm0.010$ for the effective temperatures of the primary component. 
They found that their fitting by the VR and DSED isochrones provides similar results.
These estimates of the cluster parameters agree with ours.

The mass-radius relation for V54 is fitted with the VR isochrones by \citet{vandenberg2018} and discussed by \citet{vandenberg2022}.
Analyzing also RR~Lyrae variables in agreement with the distance obtained from MS fits to local subdwarfs and modeling the cluster HB populations, 
\citet{vandenberg2022} conclude that NGC\,6809 has $(m-M)_V=13.95\pm0.05$, [Fe/H]$=-1.85\pm0.1$, [O/Fe]$=0.5\pm0.1$, $0.25<Y<0.27$, and an age of about $12.9\pm0.8$ Gyr.
All of these estimates agree with ours.

Fig.~\ref{ngc6809_v54} presents our fitting of V54 with the DSED and BaSTI isochrones for various [Fe/H], $Y$, and age. 
A better fit is seen for higher age or lower [Fe/H] or higher $Y$. The latter, about $Y=0.26$, is most fruitful to obtain the best fit.

\begin{table*}
\def\baselinestretch{1}\normalsize\normalsize
\caption[]{The relative estimates for the obtained [Fe/H] (dex), age (Gyr), distance (kpc), and $E(B-V)$ (mag) in the sense `NGC\,6809 minus NGC\,6397'. 
All the values are shown with uncertainty of one measurement.
The right column presents the differences of the parameters in the sense `NGC\,6809 minus NGC\,6397' derived from all optical CMDs. 
}
\label{relative}
\[
\begin{tabular}{lcccc}
\hline
\noalign{\smallskip}
                     &  DSED            &  BaSTI          &  Mean value       & All optical CMDs \\
\hline
\noalign{\smallskip}
[Fe/H]               & $0.04\pm0.16$    & $0.09\pm0.12$   & $0.06\pm0.14$     & $0.06$  \\
Age                  & $0.00\pm0.54$    & $0.13\pm0.44$   & $0.06\pm0.48$     & $0.04$  \\
Distance             & $2.76\pm0.09$    & $2.73\pm0.05$   & $2.75\pm0.08$     & $2.79$  \\
$E(B-V)$             & $-0.055\pm0.017$ & $-0.059\pm0.13$ & $-0.057\pm0.015$  & $-0.060$  \\
\hline
\end{tabular}
\]
\end{table*}

\subsection{Relative estimates}
\label{relative}

Similar to \citetalias{ngc6362}, we consider the relative estimates for the cluster parameters separately derived for each model.
Systematic errors of the models must be canceled out in such relative estimates. 
Therefore, the relative estimates may be more accurate than the absolute ones.

We use 8 independent CMDs of 5 twin data sets with accurate photometry in optical filters in order to derive relative estimates for the cluster parameters:
(i) $F438W-F606W$ and (ii) $F606W-F814W$ from \citetalias{nardiello2018}, (iii) $F475W-F814W$ from \citetalias{simioni2018}, (iv) $B-V$ and (v) $V-I$ from \citetalias{stetson2019}, 
(vi) $G_\mathrm{BP}-G_\mathrm{RP}$ from {\it Gaia} DR3, (vii) $g_\mathrm{SMSS}-r_\mathrm{SMSS}$, and (viii) $r_\mathrm{SMSS}-z_\mathrm{SMSS}$ from SMSS. 
Table~\ref{relative} presents the relative estimates. The models are consistent in them, i.e. the distribution of the combined sample of the DSED and BaSTI 
relative estimates for each parameter is nearly Gaussian and each uncertainty of the combined sample agrees with those of the models. 
This confirms that, indeed, systematic errors of the models are canceled out in the relative estimates.

The final uncertainties of the relative estimates are the standard deviations from Table~\ref{relative} divided by the square root of the number of the CMDs and models used
(8 CMDs by 2 models).
The relative estimates show that NGC\,6809 is $2.75\pm0.02$ kpc further, $\Delta E(B-V)=0.057\pm0.004$ less reddened, $0.06\pm0.12$ Gyr older (i.e. of nearly the same age), 
and with $0.06\pm0.03$ dex higher [Fe/H] (i.e. of nearly the same metallicity) than NGC\,6397. 
For comparison, the right column of Table~\ref{relative} presents the absolute differences of the parameters in the sense `NGC\,6809 minus NGC\,6397' derived from optical CMDs,
as described in Sect.~\ref{metal}--\ref{distance}. A good agreement between the relative estimates and absolute differences is evident.

\subsection{HB morphology difference}
\label{hb}

NGC\,6397 and NGC\,6809 show a considerable HB morphology difference (e.g. see Figs~\ref{gaia}, \ref{hst}, \ref{stetson}, \ref{stetson_details}, 
\ref{gaiavista}, and \ref{hst_336_438}) despite nearly the same metallicity and age, which are usually considered as the first and second 
parameters to explain such a difference (see \citetalias{ngc6362} and references therein).
Moreover, these clusters have similar low helium enrichments.
Therefore, we should describe this HB difference by another parameter, other than metallicity, age, or helium enrichment.

The HB morphology can be presented as the HB types (see \citealt{lee1994} for definition) of these clusters ($0.98$ for NGC\,6397 and $0.87$ for NGC\,6809 from \citealt{mackey2005} 
with an uncertainty about $\pm0.1$ from \citealt{torelli2019}), 
or as their median colour difference between the HB and RGB ($\Delta (V-I)=0.944\pm0.012$ for NGC\,6397 and $0.906\pm0.021$ for NGC\,6809 from \citealt{dotter2010}), or
an alternative HB morphology index, $\tau_{HB}=8.29\pm0.17$ for NGC\,6397 versus $6.59\pm0.21$ for NGC\,6809, which is introduced by \citet{torelli2019}.\footnote{$\tau_{HB}$ is 
calculated from cumulative number distributions along the HB in the $I$ magnitude and $V-I$ color. This index varies between 0 and 14 for the most red and blue HB.}
All these characteristics represent the HBs of NGC\,6397 and NGC\,6809 as rather blue and similar.
Hence, these indices seem to be an incomplete description of the observed significant HB morphology difference of these clusters.

Given $-2.0<$[Fe/H]$<-1.7$, age about 12.5--13.5 Gyr, and $Y\approx0.25$ for both NGC\,6397 and NGC\,6809, and taking into account the stochastic nature 
of the mass loss before the HB, e.g. described by the BaSTI ZAHB predictions (see Sect.~\ref{iso}), 
one obtains a realistic \emph{possible} scatter of the HB stars within a wide range of masses, at least $0.50-0.78$ $M_{\sun}$ for our GCs.
This \emph{possible} scatter should be compared to the \emph{observed} scatter. The latter is estimated from our fitting of the observed colour distribution of the HB and AGB 
stars by the BaSTI isochrones:
$0.63-0.67$ and $0.58-0.68$ $M_{\sun}$ for NGC\,6397 and NGC\,6809, respectively.\footnote{All stars in the middle part of the NGC\,6809's HB far from the AGB
appear RR~Lyrae variables after our star-by-star inspection.}
These estimates are obtained consistently, within $\pm0.02$ $M_{\sun}$, for all CMDs with many HB stars.
These estimates show a good agreement with those modeled by \citet{gratton2010}: 
$0.625-0.661$ and $0.645-0.682$ from {\it HST} and ground-based observations of NGC\,6397, respectively, while $0.61-0.70$ from ground-based observations of NGC\,6809.
Moreover, our mass estimates for the HB stars of NGC\,6397 agree with the estimate $0.64-0.66$ $M_{\sun}$ obtained by \citet{ahumada2021} from their modeling of the HB blue tail
with mass loss at the RGB.

Thus, the main HB morphology difference between these clusters seems to be a narrower HB mass range of NGC\,6397 w.r.t. NGC\,6809.
Note that the wider mass range of the HB stars in NGC\,6809 means a larger amount of stars evolved from them, e.g. of RR~Lyrae variables and red AGB stars.
The latter is seen in our CMDs: NGC\,6397 has no RR~Lyrae, while there are several in NGC\,6809; NGC\,6397 contains twice fewer number of the AGB stars on the red side of 
the RR~Lyrae gap.
Yet, the difference in the highest mass is rather small (0.67 versus 0.68 $M_{\sun}$). 
Hence, first of all, the desired HB parameter must explain the existence of the bluest HB stars of $0.58-0.63$ $M_{\sun}$ in NGC\,6809, but not in NGC\,6397.
Note that these stars make up the HB blue tail 
(i.e. the bluest part of the HB on the blue side of the HB knee), which is observed for NGC\,6809, but not for NGC\,6397.
Hence, a natural explanation for the HB morphology difference between these clusters is that either NGC\,6397 has lost or NGC\,6809 has acquired the bluest HB stars.
The desired parameter may be related to a peculiar evolution of these stars, i.e. with an extreme mass-loss, segregation of stellar masses due to cluster evolution, 
or other peculiar processes.

This loss or acquisition of low mass HB stars may relate to the fact that NGC\,6397 is a core-collapse cluster, i.e. the one with a highly compact, bright core, 
with a surface brightness constantly increasing towards the cluster centre.
In contrast, NGC\,6809, without core collapse, has a very low central concentration of stars and a roughly flat surface brightness of the cluster core.
The core collapse relates to the mass segregation during cluster evolution, when massive stars tend to clump in the cluster centre, while less massive ones populate the outskirts, 
sometimes escaping the cluster \citep{martinazzi2014}, as well as to the increase of dynamical interactions among stars in the dense core of post-core-collapse cluster \citep{meylan1997}.
The current mass of NGC\,6397 is just 10 per cent of its initial mass \citep{dieball2017}. 
The HB stars are the least centrally concentrated population and absent in the central area of the core of NGC\,6397 \citep{dieball2017}.
Hence, most low-mass HB stars may be lost in the dynamical evolution and mass segregation of NGC\,6397.
Yet, the additional parameter (after metallicity, age and helium enrichment) is still an issue. Our study may provide sufficient input data to solve it.

\section{Conclusions}
\label{conclusions}

This study continues the series of \citetalias{ngc5904}, \citetalias{ngc6205}, \citetalias{ngc288}, and \citetalias{ngc6362} in the estimation of key parameters
of Galactic globular clusters via fitting theoretical isochrones to observed multiband photometry.
We have analyzed the low metallicity pair NGC\,6397 and NGC\,6809 (Messier 55) with similar metallicity, age, helium enrichment, and extinction.
The cluster members have been carefully selected through {\it HST} and {\it Gaia} DR3 proper motions and parallaxes.
Accordingly, we provided the lists of reliable members of the clusters, their median parallax ($0.416\pm0.010$ and $0.203\pm0.010$ mas for NGC\,6397 and NGC\,6809, respectively),
and systemic proper motions with their total 
(systematic plus statistic) uncertainties in mas\,yr$^{-1}$:
$$\mu_{\alpha}\cos(\delta)=3.26\pm0.02,\; \mu_{\delta}=-17.66\pm0.02$$
$$\mu_{\alpha}\cos(\delta)=-3.43\pm0.02,\; \mu_{\delta}=-9.31\pm0.02$$
for NGC\,6397 and NGC\,6809, respectively.

We employed the photometry in 32 and 23 filters for NGC\,6397 and NGC\,6809, respectively, from the {\it HST}, {\it Gaia} DR3, SMSS DR3, 2MASS, VISTA VHS DR5, unWISE, and other data sets.
These filters span a wide wavelength range from the UV to mid-IR, namely from about 230\,nm to 4060\,nm. As in our previous studies, we cross-identified some data sets with each other.
As a result, we could (i) estimate systematic differences between the data sets and
(ii) use the 2MASS, VISTA and unWISE photometry with a very low extinction for determination of the extinction in all other filters and 
drawing of empirical extinction laws.

We fitted the data by the DSED and BaSTI theoretical models of stellar evolution for [$\alpha$/Fe]$=0.4$ with nearly primordial helium abundance $Y\approx0.25$.
As a result, we obtained [Fe/H], reddening, age, and distance as the parameters.
BaSTI provides metallicity $\Delta$[Fe/H]$\approx0.06$ dex systematically higher than DSED and reddening $\Delta E(B-V)\approx0.01$ mag systematically lower than DSED.

An important result of this study is the agreed parameters of NGC\,6397 and NGC\,6809 derived from successful fitting of two recent isochrone sets to all recent photometric data sets, 
most of which have never been fitted before. 
To derive reddening, age, and distance, we use 19 and 14 independent CMDs, while 16 and 10 ones to derive [Fe/H] for NGC\,6397 and NGC\,6809, respectively.

The following estimates were obtained for NGC\,6397 and NGC\,6809, respectively:
metallicities [Fe/H]$=-1.84\pm0.02\pm0.1$ and $-1.78\pm0.02\pm0.1$ (statistic and systematic uncertainties);
distances $2.45\pm0.02\pm0.06$ and $5.24\pm0.02\pm0.18$ kpc;
distance moduli $(m-M)_0=11.95\pm0.01\pm0.05$ and $13.60\pm0.01\pm0.07$ mag;
apparent $V$-band distance moduli $(m-M)_\mathrm{V}=12.54\pm0.03\pm0.06$ and $13.97\pm0.03\pm0.08$ mag;
ages $12.9\pm0.1\pm0.8$ and $13.0\pm0.1\pm0.8$ Gyr;
reddenings $E(B-V)=0.178\pm0.006\pm0.01$ and $0.118\pm0.004\pm0.01$ mag; 
extinctions $A_\mathrm{V}=0.59\pm0.01\pm0.02$ and $0.37\pm0.01\pm0.04$ mag; and
extinction-to-reddening ratio $R_\mathrm{V}=3.32^{+0.32}_{-0.28}$ and $3.16^{+0.66}_{-0.56}$.
These estimates agree with most estimates from the literature, while disapprove other estimates.
For example, after our [Fe/H] estimates, higher [Fe/H] estimates by \citet{meszaros2020} seem to be preferred over the lower ones by \citet{carretta2009}. 

There are pairs of similar data sets for the clusters, which are obtained with the same telescope and/or processed within the same pipeline.
We used these data sets to derive very precise relative estimates for the parameters.
NGC\,6809 appears $2.75\pm0.02$ kpc further, $\Delta E(B-V)=0.057\pm0.004$ less reddened, $0.06\pm0.12$ Gyr older (i.e. of the same age), and with $0.06\pm0.03$ dex higher [Fe/H]
(i.e. of the same metallicity) than NGC\,6397.

Despite nearly the same metallicity, age, and helium enrichment, these clusters show a considerable HB morphology difference, which must therefore be described by another parameter.
Primarily, this parameter must explain the existence of the least massive HB stars of the blue tail (0.58--0.63 solar mass) only in NGC\,6809.
Probably such stars have been lost by the core-collapse cluster NGC\,6397 in its dynamical evolution and mass segregation, unlike NGC\,6809, which has a very low central concentration.

\section*{Acknowledgements}

We acknowledge financial support from the Russian Science Foundation (grant no. 20--72--10052).

We thank the anonymous reviewer for useful comments.
We thank 
Armando Arellano Ferro for providing and discussion of the $VI$ photometry,
Eugenio Carretta for discussion of cluster metallicity,
Santi Cassisi for providing the valuable BaSTI isochrones and his useful comments,
Aaron Dotter for his comments on DSED,
Christopher Onken, Taisia Rahmatulina and Sergey Antonov for their help to access the SkyMapper Southern Sky Survey DR3,
Peter Stetson for providing the valuable $UBVRI$ photometry.

This work has made use of BaSTI and DSED web tools;
Filtergraph \citep{filtergraph}, an online data visualization tool developed at Vanderbilt University through the Vanderbilt Initiative in 
Data-intensive Astrophysics (VIDA) and the Frist Center for Autism and Innovation (FCAI, \url{https://filtergraph.com});
the resources of the Centre de Donn\'ees astronomiques de Strasbourg, Strasbourg, France (\url{http://cds.u-strasbg.fr}), including the SIMBAD database, 
the VizieR catalogue access tool \citep{vizier} and the X-Match service;
observations made with the NASA/ESA {\it Hubble Space Telescope};
data products from the {\it Wide-field Infrared Survey Explorer}, which is a joint project of the University of California, Los Angeles, and the Jet Propulsion 
Laboratory/California Institute of Technology;
data products from the Two Micron All Sky Survey, which is a joint project of the University of Massachusetts and the Infrared Processing and Analysis Center/California 
Institute of Technology, funded by the National Aeronautics and Space Administration and the National Science Foundation;
data products from the Pan-STARRS Surveys (PS1);
data from the European Space Agency (ESA) mission {\it Gaia} (\url{https://www.cosmos.esa.int/gaia}), processed by the {\it Gaia} Data Processing and Analysis Consortium 
(DPAC, \url{https://www.cosmos.esa.int/web/gaia/dpac/consortium}), and {\it Gaia} archive website (\url{https://archives.esac.esa.int/gaia});
data products from the SkyMapper Southern Sky Survey, SkyMapper is owned and operated by The Australian National University's Research School of Astronomy and Astrophysics,
the SkyMapper survey data were processed and provided by the SkyMapper Team at ANU, the SkyMapper node of the All-Sky Virtual Observatory (ASVO) is hosted at the National 
Computational Infrastructure (NCI).

\section*{Data availability}

The data underlying this article will be shared on reasonable request to the corresponding author.

\bsp	
\label{lastpage}
\end{document}